\documentclass{article}

\usepackage[preprint]{arxiv_preprint}

\usepackage[utf8]{inputenc} 
\usepackage[T1]{fontenc}    
\usepackage{hyperref}       
\usepackage{url}            
\usepackage{array, booktabs, subcaption, caption}       
\usepackage{amsfonts}       
\usepackage{amsmath, amssymb, tikz-cd}        
\usepackage{nicefrac}       
\usepackage[expansion=false,patch=none]{microtype}      
\usepackage{xcolor}         
\usepackage{tcolorbox}      
\usepackage{placeins}     
\usepackage{multirow}

\makeatletter
\renewcommand{\@notice}{}           
\makeatother

\title{LeanSearch v2: Global Premise Retrieval for Lean 4 Theorem Proving}

\author{%
  \parbox{0.95\textwidth}{\centering
  Guoxiong Gao\textsuperscript{1,2} \quad Zeming Sun\textsuperscript{1,3} \quad Jiedong Jiang\textsuperscript{4} \quad Yutong Wang\textsuperscript{1,2} \\[2pt]
  Jingda Xu\textsuperscript{2} \quad Peihao Wu\textsuperscript{2} \quad Bryan Dai\textsuperscript{2,\,$\dagger$} \quad Bin Dong\textsuperscript{5,6,7,8,\,$\dagger$} \\[6pt]
  \normalfont\small
  \textsuperscript{1}School of Mathematical Sciences, Peking University \\
  \textsuperscript{2}IQuest Research \\
  \textsuperscript{3}Research Institute for Mathematical Sciences, Kyoto University \\
  \textsuperscript{4}Westlake Institute for Advanced Study, Westlake University \\
  \textsuperscript{5}Beijing International Center for Mathematical Research and the New Cornerstone Science Laboratory, Peking University \\
  \textsuperscript{6}Center for Machine Learning Research, Peking University \\
  \textsuperscript{7}Center for Intelligent Computing, Great Bay Institute for Advanced Study, Great Bay University \\
  \textsuperscript{8}Zhongguancun Academy \\[4pt]
  \textsuperscript{$\dagger$}Corresponding authors: \texttt{cbdai@iquestlab.com}, \texttt{dongbin@math.pku.edu.cn}
  }
}

\begin{document}

\maketitle

\begin{abstract}
Proving theorems in Lean~4 often requires identifying a scattered set of library lemmas whose joint use enables a concise proof---a task we call \emph{global premise retrieval}. Existing tools address adjacent problems: semantic search engines find individual declarations matching a query, while premise-selection systems predict useful lemmas one tactic step at a time. Neither recovers the full premise set an entire theorem requires. We present LeanSearch~v2, a two-mode retrieval system for this task. Its standard mode applies a hierarchy-informalized Mathlib corpus with an embedding--reranker pipeline, achieving state-of-the-art single-query retrieval without domain-specific fine-tuning (nDCG@10 of $0.62$ vs.\ $0.53$ for the next-best system). Its reasoning mode builds on standard mode as its retrieval substrate, targeting global premise retrieval through iterative sketch-retrieve-reflect cycles. On a 69-query benchmark of research-level Mathlib theorems, reasoning mode recovers $46.1\%$ of ground-truth premise groups within 10 retrieved candidates, outperforming strong reasoning retrieval systems ($38.0\%$) and premise-selection baselines ($9.3\%$) on the same benchmark. In a controlled downstream evaluation with a fixed prover loop, replacing alternative retrievers with LeanSearch~v2 yields the highest proof success ($20\%$ vs.\ $16\%$ for the next-best system and $4\%$ without retrieval), confirming that retrieval quality propagates to proof generation. We have open-sourced all code, data, and benchmarks.\footnote{Code and data are available at \url{https://github.com/frenzymath/LeanSearch-v2}.}\footnote{The standard mode is publicly available with API access at \url{https://leansearch.net/}.}
\end{abstract}

\section{Introduction}
\label{sec:intro}

Rapid advances in Lean~4 and its central library Mathlib~\citep{mathlib} have made formalization an increasingly accepted practice in the mathematical community, with the library now containing over one hundred thousand verified declarations spanning algebra, analysis, topology, and number theory. AI systems have further accelerated this trend, demonstrating growing capability in generating formal proofs~\citep{alphaproof,deepseekproverv2,aristotle}. A central challenge in working within this ecosystem is leveraging what Mathlib already contains. Locating the right library lemma carries two complementary benefits: a well-chosen result can collapse a lengthy derivation into a single tactic invocation, and reusing existing results is the community's explicit coding standard. Mathlib prioritizes generality and actively discourages code duplication~\citep{mathlibcontrib}; indeed, ``being able to find the library theorems you need constitutes an important part of formalization''~\citep{mathliblean}. This norm has direct consequences for automated provers: without effective retrieval, an AI system must repeatedly prove local versions of lemmas that already exist in the library for each problem it encounters, wasting computation and increasing proof difficulty. Yet finding the right lemmas is hard, not primarily because the library is large, though its scale compounds the problem, but because the connections require mathematical reasoning. Effectively applying a powerful theorem often demands reshaping the proof state to match its interface, generalizing or transforming the current goal in ways that temporarily complicate it, only to yield a far simpler overall proof by routing through a result in a seemingly unrelated area. The relevant lemmas are linked to the problem not by shared vocabulary but by the logical architecture of a proof strategy. Retrieving the \emph{set} of supporting lemmas on which a complete proof should rest is a task we term \emph{global premise retrieval}: identifying which library results are essential to the theorem's overall proof strategy, considering the needs of the entire proof rather than what simplifies any single intermediate state.

\begin{figure}[ht]
\centering
\begin{tcolorbox}[colback=gray!5, colframe=gray!60, title={\small\textbf{Example: Global Premise Retrieval}}, fonttitle=\small, boxrule=0.4pt, left=4pt, right=4pt, top=2pt, bottom=2pt]
\small
\textbf{Problem.} Prove that $\frac{x^p - 1}{x - 1}$ is irreducible in $\mathbb{Z}[x]$ for every prime $p$.\\[3pt]
\textbf{Supporting premises (from 3 distant Mathlib modules):}\\[2pt]
\begin{tabular}{@{}>{\raggedright\arraybackslash}p{0.4\linewidth}>{\raggedright\arraybackslash}p{0.55\linewidth}@{}}
\toprule
\textbf{Lemma} & \textbf{Statement / Role} \\
\midrule
\texttt{geom\_sum\_mul} \newline {\scriptsize\color{gray}Algebra.Ring.GeomSum} & $(\sum_{i=0}^{n-1} x^i)(x{-}1) = x^n {-} 1$\newline Establishes the algebraic identity. \\[3pt]
\texttt{Polynomial.cyclotomic\_prime} \newline {\scriptsize\color{gray}RingTheory.Polynomial.Cyclotomic.Basic} & $\Phi_p(x) = \sum_{i=0}^{p-1} x^i$\newline Identifies the quotient as the $p$-th cyclotomic polynomial.\\[3pt]
\texttt{Polynomial.cyclotomic.\allowbreak irreducible} \newline {\scriptsize\color{gray}RingTheory.Polynomial.Cyclotomic.Roots} & $\Phi_n$ is irreducible over $\mathbb{Z}$ for $n > 0$.\newline Concludes the proof in one step.\\
\bottomrule
\end{tabular}\\[4pt]
{\footnotesize\textit{Note.} The problem statement shares no vocabulary with these premises (``cyclotomic,'' ``geometric sum,'' ``$\Phi$'' all absent). Identifying them requires reasoning about proof strategy, not lexical matching; the three lemmas chain together to close the proof, and missing any one leaves the argument incomplete.}
\end{tcolorbox}
\caption{A global premise retrieval instance requiring three lemmas from distant Mathlib modules.}
\label{fig:example}
\end{figure}

Figure~\ref{fig:example} illustrates why this task is difficult. The classical proof of irreducibility of $(x^p{-}1)/(x{-}1)$ proceeds by substitution and Eisenstein's criterion, but a more efficient proof in Lean bypasses this re-derivation entirely by routing through existing library results on cyclotomic polynomials, resolving the goal in one step once the algebraic equivalence is established. Recognizing that such a result exists is the first challenge, since the problem statement shares no vocabulary with the cyclotomic machinery. The second challenge is pinpointing the three specific lemmas that chain together to close the proof; missing any one leaves the argument incomplete. In graduate-level mathematics and beyond, this pattern is extremely common: a proof relies on a handful of supporting results dispersed across the library, connected not by shared terminology but by the logical architecture of the proof strategy.

Two active research lines build retrieval tools for mathematics, yet neither addresses global premise retrieval. The \emph{premise selection} line~\citep{reprover,leanstatesearch,leansearchps,leanhammer} retrieves premises for a given proof state, improving tactic-level proving but operating at per-state granularity without considering what an entire proof requires. The \emph{reasoning retrieval} line, catalyzed by the BRIGHT benchmark~\citep{bright}, advances retrieval for queries requiring multi-step inference through reasoning-aware encoders, multi-stage pipelines, and LLM-based reranking~\citep{reasonir,diver,inf-x}. However, current benchmarks evaluate each retrieved document's relevance independently; these systems have not yet addressed the regime where multiple premises from different mathematical areas must jointly support a proof, nor the formal-language setting where problem statements may be deliberately reshaped to match a library theorem's interface. Neither line is designed to recover a \emph{logically coherent set} of premises for an entire theorem, where the elements are connected by the internal logic of a proof strategy rather than merely co-relevant to a query.

We present \textbf{LeanSearch~v2}, a retrieval system designed for this task. Its \emph{standard mode} serves as the foundation: it builds a hierarchy-informalized Mathlib corpus and retrieves against it with a two-stage embedding--reranker pipeline, without domain-specific fine-tuning. Its \emph{reasoning mode} builds on standard mode as its retrieval substrate to target global premise retrieval directly: given a theorem statement, it generates a proof sketch, retrieves candidates for each proof step via standard mode, filters and judges the results, and iteratively revises the sketch until a coherent supporting-lemma set emerges. We evaluate on two new benchmarks: \textbf{MathlibQR} (Mathlib Queries), a $200$-declaration, $946$-query evaluation for standard-mode search, and \textbf{MathlibMPR} (Mathlib Merged Pull Requests), a $69$-theorem benchmark with expert-annotated premise groups (one to eight per theorem) for global premise retrieval. On MathlibQR, standard mode achieves nDCG@$10$ of $0.62$ against $0.53$ for the next-best system. On MathlibMPR, reasoning mode recovers $46.1\%$ of ground-truth premise groups within $10$ retrieved candidates, outperforming both reasoning retrievers (INF-X~\citep{inf-x}, ReasonIR~\citep{reasonir}, DIVER~\citep{diver}) and premise-selection baselines (ReProver~\citep{reprover}, LeanStateSearch~\citep{leanstatesearch}, LeanPremise~\citep{leanhammer}) on the same benchmark. A controlled downstream experiment further confirms that this retrieval advantage propagates to proof success when the prover loop is held fixed.

Although we evaluate the downstream effect of plugging our retriever into a prover system (\S\ref{sec:exp-prove}), we deliberately position this work in the retrieval-research stratum rather than the proof-generation stratum. Strong agent systems~\citep{seedprover,numinaagent,aristotle,archon} have shown a capable agent can compensate for a weaker retriever through repeated query reformulations, or bypass retrieval entirely by proving needed lemmas locally. However, closing the gap between a proof that merely compiles and an efficient, idiomatic proof still depends on identifying and using existing results—a core challenge that makes global premise retrieval a meaningful task in its own right.

\paragraph{Contributions.}
\begin{itemize}
\setlength{\itemsep}{0pt}
\item We formulate \emph{global premise retrieval}, the task of recovering the logically coherent set of library lemmas that an entire theorem's proof requires, and construct two expert-curated Lean~4 benchmarks: MathlibQR ($200$ declarations, $946$ queries) for single-query search, and MathlibMPR ($69$ theorems with structured premise-group annotations and alternative proof routings) for set-valued premise retrieval.
\item We present LeanSearch~v2, a staged retrieval system whose standard mode achieves state-of-the-art Mathlib search without domain-specific fine-tuning, and whose reasoning mode builds on it through iterative sketch-retrieve-reflect cycles to target global premise retrieval directly, outperforming both reasoning-intensive retrievers and premise-selection baselines on our benchmarks. A controlled downstream experiment further confirms that retrieval quality propagates to end-to-end proof success.
\item To our knowledge, this is the first work to bring reasoning-retrieval techniques to the Lean premise retrieval problem.
\end{itemize}

\section{Related Work}

\subsection{Reasoning Retrieval}

Recent benchmarks formalize retrieval tasks that require reasoning beyond lexical matching~\citep{naturalproofs,bright,mirb,crumb,mmbright}. \citet{naturalproofs} introduced mathematical reference retrieval in informal text. BRIGHT~\citep{bright} subsequently defined \emph{reasoning-intensive retrieval}, revealing a large accuracy drop when standard dense retrievers face queries requiring multi-step inference. Follow-up methods address this through reasoning-aware encoders~\citep{reasonir,rader}, multi-stage pipelines~\citep{diver}, and LLM-based reranking~\citep{rank1,grouprank}. These benchmarks and methods target informal-text corpora such as web documents and scientific articles, evaluating each retrieved document's relevance independently.

\subsection{Retrieval on Mathlib}

\paragraph{Semantic search engines.} Several tools support natural-language search over Lean~4's Mathlib. Moogle\footnote{\url{https://www.moogle.ai}} and Loogle\footnote{\url{https://loogle.lean-lang.org/}} provide semantic and syntactic search respectively. LeanSearch~\citep{leansearchv1} applies off-the-shelf embeddings to a bilingual formal--informal corpus. LeanExplore~\citep{leanexplore} combines dense, sparse, and graph-based signals. Lean Finder~\citep{leanfinder} fine-tunes on synthesized user-intent queries. Each system returns a ranked list for a single natural-language query.

\paragraph{Premise selection.} A separate line retrieves premises conditioned on a local proof state. ReProver~\citep{reprover} introduced dense retrieval over state--premise pairs in Lean. Subsequent work refines the paradigm with improved negative mining~\citep{leanstatesearch}, integration into end-to-end provers~\citep{leansearchps}, automated reasoning backends~\citep{leanhammer}, and graph-augmented embeddings~\citep{gnnreprover}. These systems focus on finding premises that advance the current local proof state.

\section{Methodology}

LeanSearch~v2 operates in two modes.
\emph{Standard mode} (\S\ref{subsec:standard}) retrieves individual Mathlib declarations from a natural-language query.
It serves as both a standalone search engine and the retrieval substrate for the second mode.
\emph{Reasoning mode} (\S\ref{subsec:reasoning}) targets global premise retrieval: given a formal theorem statement, it iteratively sketches proof strategies and queries standard mode, assembling the set of library premises needed for a concise proof.

\subsection{Standard mode}\label{subsec:standard}

\begin{figure}[ht]
    \centering
    \includegraphics[width=\textwidth]{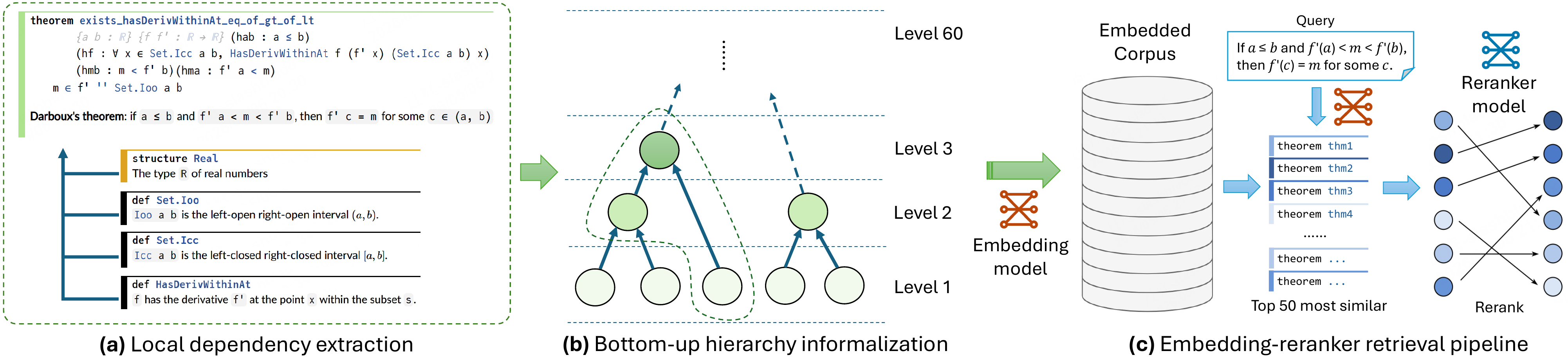}
    \caption{Standard mode pipeline. \textbf{(a)}~Jixia extracts each declaration and its dependencies from Mathlib. \textbf{(b)}~Declarations are informalized bottom-up so that each description can draw on already-informalized dependencies. \textbf{(c)}~Queries and corpus items are embedded for cosine-similarity retrieval; a reranker refines the top candidates.}
    \label{fig:standard-mode}
\end{figure}

Standard mode serves as both a standalone search engine and the retrieval substrate for reasoning mode (Figure~\ref{fig:standard-mode}).
It comprises two components: an informalized Mathlib corpus and an embedding--reranker retrieval pipeline.

\paragraph{Informalized corpus preparation.}

The retrieval corpus covers all extractable declarations from Mathlib v4.28.0-rc1 (definitions, theorems, type classes, instances, and other named constants), each paired with a concise natural-language description of its mathematical meaning.

\begin{figure}[ht]
    \centering
    \includegraphics[width=1.0\linewidth]{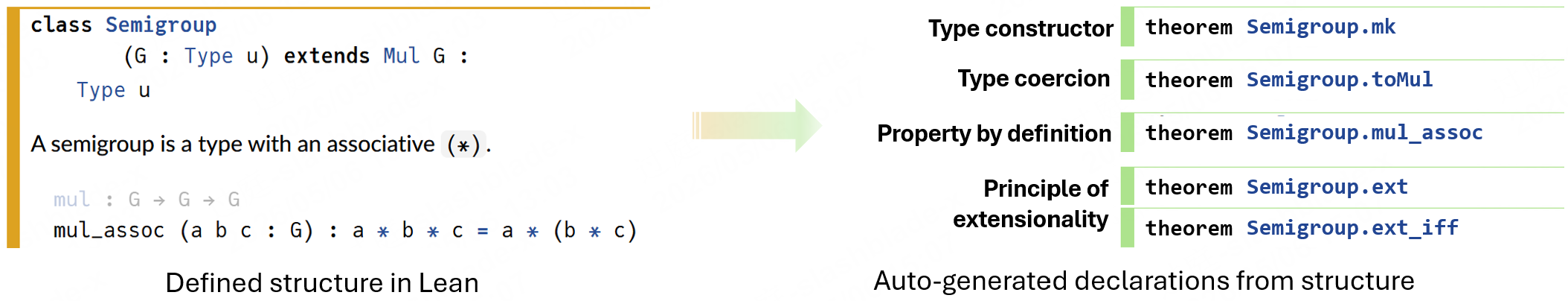}
    \caption{Hidden auto-generated declarations in Mathlib.}
    \label{fig:hidden-decls}
\end{figure}

We extract declarations using Jixia,\footnote{\url{https://github.com/frenzymath/jixia}} a publicly available, open-source, non-intrusive static analysis tool for Lean~4 (Figure~\ref{fig:standard-mode}a).
Its Declaration plugin provides source-level metadata (kind, arguments, return type, source range); its Symbol plugin records elaborated constants, their types, and the inter-declaration reference graph.
To improve extraction coverage, Jixia is extended to
(i)~extract hidden compiler-generated declarations such as auxiliary definitions produced for structures (Figure~\ref{fig:hidden-decls}),
(ii)~recover declarations from elaborated terms rather than surface source alone, and
(iii)~support customized Lean runtime options so that extracted representations match the configuration used during corpus construction.

The reference graph induces a directed acyclic graph (DAG) over declarations.
We topologically sort it and informalize bottom-up (Figure~\ref{fig:standard-mode}b): each call receives the already-informalized descriptions of the target's dependencies as context, so that descriptions are grounded in mathematical concepts rather than opaque formal identifiers; this dependency-aware informalization strategy is known to produce more grounded descriptions than flat, context-free informalization~\citep{herald}.
Informalization uses Qwen3-32B~\citep{qwen3} with task-specific prompts; failures are reprocessed with Gemini~2.5~Pro~\citep{gemini25} as a fallback.
Each corpus item stores declaration kind, fully qualified name, type signature, value (when available), source location, dependencies, and the informalized description.

\paragraph{Embedding and reranker pipeline.}

Each corpus item is encoded into a fixed-dimensional vector using Qwen3-Embedding-8B~\citep{qwen3embedding}, a general-purpose embedding model that ranks among the strongest on MTEB~\citep{mteb}, the standard benchmark suite for text embedding quality, without domain-specific fine-tuning (Figure~\ref{fig:standard-mode}c).
For each item we carefully compose a structured passage from a tuned template that combines a task instruction with well-formatted formal and informal information, including the declaration kind, type signature, informalized description, and, for definitions, classes, and instances, the \texttt{value} field, whose content carries essential semantic meaning (unlike theorem values, which are proof terms).
The passages are embedded and indexed for cosine-similarity retrieval.

The top-50 candidates are reranked by Qwen3-Reranker-8B~\citep{qwen3embedding}, which scores each query-candidate pair by predicting a binary relevance token and using $P(\texttt{yes})$ as the score.
The reranker receives a kind-aware instruction with special handling for definition-type declarations, aiming to compensate for the retrieval difficulty on definitions that prior approaches~\citep{leansearchv1}, which did not distinguish declaration kinds, consistently exhibited.
An ablation showing the impact of this presentation design on definition retrieval is in Appendix~\ref{app:search-ablation}.
The reranked list constitutes standard mode's final output.

No task-specific training is performed: neither the embedder nor the reranker is fine-tuned on Mathlib data.
Domain knowledge enters through informalization and kind-aware prompting, keeping the system straightforward to maintain as Mathlib evolves, and to transfer to other rapidly growing AI-generated formal libraries whose style and conventions may differ substantially from Mathlib.

\subsection{Reasoning mode}\label{subsec:reasoning}

\begin{figure}[ht]
    \centering
    \includegraphics[width=\textwidth]{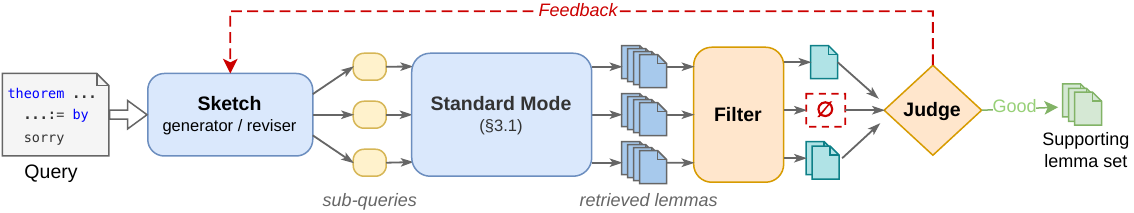}
    \caption{Reasoning mode pipeline. The sketch generator decomposes a target theorem into sub-queries. Each sub-query is sent to standard mode (\S\ref{subsec:standard}), and a filter vets the retrieved candidates, potentially returning an empty set ($\varnothing$). The judge inspects all filtered outputs and either accepts the supporting lemma set or sends structured feedback (red dashed arrow) to the sketch reviser for another iteration.} 
    \label{fig:reasoning-mode}
\end{figure}

Reasoning mode targets global premise retrieval through an iterative sketch-retrieve-reflect loop built on top of standard mode (Figure~\ref{fig:reasoning-mode}).
Given a target theorem, the loop decomposes it into sub-queries via a sketch generator, retrieves candidates for each sub-query through standard mode (\S\ref{subsec:standard}), filters the results, and submits the filtered map to a feasibility judge that either accepts the supporting lemma set or routes structured feedback back to a sketch reviser for another iteration.
Standard mode is treated as a black-box retrieval service: reasoning mode only issues natural-language queries against it and consumes ranked results, never reaching into the embedder or reranker.
This separation lets the loop reason about \emph{retrieval signals}, including the absence of useful results for a given query, rather than about embeddings or scores, and keeps each mode independently improvable.
We use Claude Sonnet 4.5~\citep{sonnet45} for sketch generation, judge, and revision, and Kimi K2 Instruct~\citep{kimik2} for the filter.

\paragraph{Sketch generation and revision.}

Given a target theorem (its formal statement and an informal description), the sketch generator proposes a step-by-step proof outline (Figure~\ref{fig:reasoning-mode}, left).
Each step contains a natural-language description of the mathematical move, any auxiliary context it relies on, and a retrieval query phrased in the style that standard mode expects.
The generator is not asked to write Lean code or name specific Mathlib lemmas; its task is to commit to a proof strategy at the granularity of \emph{which results should exist and how each contributes}, leaving lemma identification to the retrieval substrate.
The sketch reviser shares this interface but additionally consumes the previously rejected sketch and the judge's feedback.

\paragraph{Document filter.}

For each sub-query in the current sketch, standard mode returns a ranked list of candidate declarations (Figure~\ref{fig:reasoning-mode}, center).
The filter LLM inspects each candidate's metadata (kind, type signature, value, informalized statement) alongside the query and independently labels it as relevant or not.
Crucially, the filter may return an empty set: when no candidate is genuinely useful for the planned step, none is kept.
This empty signal allows the judge to distinguish ``the retriever found support for this step'' from ``the retriever found nothing useful here,'' two situations that demand different revisions and that a top-$k$ rule would conflate.

\paragraph{Feasibility judge and reflection loop.}

After all per-step filtered outputs are collected, the judge inspects the full filtered map and decides whether the sketch as a whole is supportable by the library (Figure~\ref{fig:reasoning-mode}, right).
Judgment is binary: if supportable, the loop terminates and the per-query filtered lists are pooled with an nDCG-style rank discount to form reasoning mode's final output (formula and deduplication rule in Appendix~\ref{app:reasoning-aggregation}); if unsupportable, the judge produces structured feedback identifying which steps failed and why (e.g., a missing premise, a candidate suggesting a different proof route, or an unsatisfiable type constraint), and routes control to the sketch reviser for a revised sketch that re-enters the retrieve--filter--judge pipeline.
We cap the loop at three rounds of revision past the initial sketch; if the judge has not accepted by then, we take the latest filtered output regardless. An ablation against a no-reflection variant of the loop is in Appendix~\ref{app:reflection-ablation}.
Loop-iteration statistics are reported in Appendix~\ref{app:reasoning-loop-stats}.

\section{Experiments}
\label{sec:exp}

We evaluate LeanSearch~v2 across three tasks.
\textbf{Search} (\S\ref{sec:exp-search}) evaluates standard mode (\S\ref{subsec:standard}) in the canonical Mathlib search setting: given a natural-language query, retrieve the matching declaration, scored by nDCG@\,$k$ and Recall@\,$k$.
\textbf{Global premise retrieval} (\S\ref{sec:exp-premise}) evaluates reasoning mode (\S\ref{subsec:reasoning}) on the task formulated in \S\ref{sec:intro}: given a theorem statement, retrieve the set of supporting lemmas. Benchmark construction and ground-truth annotation are detailed in \S\ref{sec:exp-premise}.
\textbf{Prove} (\S\ref{sec:exp-prove}) measures retrieval quality extrinsically: each retriever is plugged into a fixed proof-generation workflow (the \emph{simple reflection loop}, see \S\ref{sec:exp-prove}), and we report end-to-end proof success on FATE-H~\citep{fate}, a $100$-problem benchmark sitting at the graduate-level tier of the FATE algebra series (FATE-M / H / X span undergraduate to post-PhD difficulty), and on MathlibMPR-Prop, a $50$-problem subset of MathlibMPR (\S\ref{sec:exp-premise}).
The first task compares standard mode against existing Mathlib search engines, while the latter two compare reasoning mode against both reasoning-intensive retrievers and Lean-side premise-selection systems; to our knowledge no prior evaluation places these two families on a single Lean benchmark.

\subsection{Search: query-to-declaration retrieval}
\label{sec:exp-search}

The Search task measures standard mode's quality as a Mathlib search engine: given a natural-language query, recover the single matching declaration. We evaluate on
\textbf{MathlibQR} (Mathlib Queries), a benchmark we construct by sampling $200$ Mathlib
declarations across all major mathematical subjects and all major
Mathlib declaration kinds, with up to six query styles per declaration
(Lean-flavored, LaTeX, plain English, conceptual slogan, informal
nickname, and special-case instance) for $\mathbf{946}$ query rows in
total; ground-truth declarations are tagged \emph{Easy} or \emph{Hard} based on mathematical obscurity and technical complexity (rubric in Appendix~\ref{app:mathlibqr}); every query inherits its declaration's tag.
The benchmark intentionally spans diverse query styles and declaration kinds so that no system can succeed by overfitting to a single style; further examples and full curation details are provided in Appendix~\ref{app:mathlibqr}.

We compare four systems: LeanExplore~\citep{leanexplore}, a hybrid
dense / sparse / graph engine; LeanFinder~\citep{leanfinder}, a
fine-tuned dense retriever with a learned reranker; LeanSearch~v2
with its full retrieve-rerank pipeline; and a retriever-only
ablation of LeanSearch~v2 that drops the Qwen3-Reranker-8B stage,
which we report alongside the full system to expose the reranker's
marginal contribution. Because these systems target different
Mathlib snapshots (i.e., a declaration available to one may be absent in another), the main report uses the \emph{fair} subset of
MathlibQR, the $810$ query rows whose ground-truth declarations
exist in every system's snapshot, so that the comparison controls
for snapshot drift rather than retrieval quality; results on looser
subsets ($937$-row and full $946$-row) are in
Appendix~\ref{app:mathlibqr-perspectives}.

We report nDCG@\,$k$ and Recall@\,$k$ with binary single-document
relevance, and additionally an LLM-as-judge protocol following
\citet{leanexplore}: each system's top-$5$ returns are presented to
Claude Sonnet~4.5 in three independent random permutations per query,
and the judge produces a strict total ordering $1$--$4$ across the
four systems. We report mean rank (lower is better) over $2430$
judgments per system; position-bias and cross-round agreement
diagnostics, including a primacy-bias correction, are reported in
Appendix~\ref{app:llm-judge}.

\begin{table}[!t]
\centering
\caption{\textbf{Search task on MathlibQR.} Numbers are
nDCG@\,$k$ and Recall@\,$k$ over $810$ query rows on the shared
$171$-declaration corpus subset (intersection of the four systems'
Mathlib snapshots). \emph{LLM judge} reports mean rank
($\downarrow$ better, range $1$--$4$) by Claude Sonnet~4.5 over
$2430$ blinded judgments per system. Best per column in
\textbf{bold}.}
\label{tab:search-main}
\small
\setlength{\tabcolsep}{4pt}
\begin{tabular}{lcccccccc}
\toprule
& \multicolumn{3}{c}{nDCG@\,$k$} & \multicolumn{3}{c}{Recall@\,$k$} & & LLM judge \\
\cmidrule(lr){2-4}\cmidrule(lr){5-7}\cmidrule(lr){9-9}
System & 1 & 5 & 10 & 10 & 50 & 100 & & mean rank \\
\midrule
LeanExplore                          & 0.246 & 0.358 & 0.393 & 0.569 & 0.743 & 0.789 & & 3.32 \\
LeanFinder                           & 0.370 & 0.514 & 0.533 & 0.698 & 0.824 & \textbf{0.875} & & 2.87 \\
LeanSearch~v2 (retriever-only)       & 0.340 & 0.472 & 0.494 & 0.657 & 0.790 & 0.830 & & 2.18 \\
LeanSearch~v2 (rerank)               & \textbf{0.470} & \textbf{0.601} & \textbf{0.623} & \textbf{0.780} & \textbf{0.847} & 0.858 & & \textbf{1.63} \\
\bottomrule
\end{tabular}
\end{table}

LeanSearch~v2 (Table~\ref{tab:search-main}, last row) leads on most
reported metrics. The
LLM judge column corroborates this ranking, with
LeanSearch~v2 attaining a mean rank of $1.63$.
Appendices~\ref{app:search-slices} and~\ref{app:search-ablation} report per-slice breakdowns and a step-wise ablation: v2-rerank leads on nearly every per-slice cell, with the rerank step accounting for roughly $+10$ nDCG/Recall points at $k\!\le\!10$ and kind-aware prompting adding another $+2$--$3$.
\FloatBarrier

\subsection{Global premise retrieval}
\label{sec:exp-premise}

The global premise retrieval task evaluates reasoning mode (\S\ref{subsec:reasoning}) on the formulation of \S\ref{sec:intro}: given a theorem statement, recover the full set of library lemmas its proof requires. We assemble \textbf{MathlibMPR} (Mathlib Merged Pull Requests), a benchmark of $69$ theorems sourced from merged Mathlib pull requests. For each PR we extract its main statement, an informal description of the goal, and the set of premises actually invoked by its tactic proof. Ground truth is structured into \emph{premise groups} (one to eight per query, mean $2.96$), where lemmas within a group are interchangeable for a given proof step; for a subset of queries, formalization experts additionally annotate equivalent proof re-routings as \emph{alternative routings}. The benchmark's central design choice is mathematical substance: candidate PRs are filtered to retain only theorems that are not trivial restatements, special cases, or notational rewrites, so that even single-group queries reflect nontrivial use of a premise. Compared to traditional reasoning retrieval benchmarks, global premise retrieval poses significantly greater difficulty and exhibits distinct characteristics; a detailed comparison is given in Appendix~\ref{app:bright}. Full curation details and examples are provided in Appendix~\ref{app:benchmark-curation}.

We compare against two complementary baseline families. The first
is the reasoning-intensive retrieval line (INF-X-Retriever~\citep{inf-x}, ReasonIR~\citep{reasonir}, and DIVER~\citep{diver}), which we run over our hierarchy-informalized Mathlib corpus (\S\ref{subsec:standard}) so that they read each declaration in the same dependency-aware natural-language form that LeanSearch~v2 uses. All published
ablations of INF-X and DIVER are reported individually so that the
comparison includes each system at its strongest configuration. The
second is the Lean-side premise-selection line, which conditions on
the formal proof state of the target theorem (extracted through the
Lean~4 REPL; see Appendix~\ref{app:retriever-io}): ReProver~\citep{reprover},
LeanPremise~\citep{leanhammer}, and LeanStateSearch~\citep{leanstatesearch}. Reasoning
retrievers and our reasoning-mode pipeline take the informal
description $+$ formal statement as input; the premise-selection
systems take the formal proof state, matching their published query
contract. Per-system input details are in
Appendix~\ref{app:retriever-io}.

We report two top-$k$ metrics over $k \in \{5, 10, 20, 30, 50\}$.
\textbf{Recall (group)} is the fraction of ground-truth premise groups
hit (where a group counts as hit if any of its interchangeable lemmas
appears in the top-$k$), macro-averaged over queries. \textbf{Covered}
is the per-query indicator of whether some complete proof routing
(the original PR routing or an expert-annotated alternative) has \emph{every} premise group hit. 
An ablation that uses only the original PR routing as the success
criterion is reported in Appendix~\ref{app:premise-ablation}.

\begin{table}[!t]
\centering
\caption{\textbf{Global premise retrieval task on MathlibMPR.}
Numbers are percentages. \emph{Recall (group)}: fraction of
ground-truth premise groups hit by the top-$k$, macro-averaged.
\emph{Covered}: $1$ iff some complete proof routing (original or
expert-annotated alternative) has all of its premise groups hit by the
top-$k$. Top block: reasoning retrievers consuming informal
description $+$ formal statement; sub-rows give published ablations.
Bottom block: premise-selection systems consuming the extracted
proof state. Best per column in \textbf{bold}.}
\label{tab:premise-main}
\small
\setlength{\tabcolsep}{4pt}
\begin{tabular}{lccccc@{\hspace{6pt}}ccccc}
\toprule
& \multicolumn{5}{c}{Recall@\,$k$ (group)} & \multicolumn{5}{c}{Covered@\,$k$} \\
\cmidrule(lr){2-6}\cmidrule(lr){7-11}
System & 5 & 10 & 20 & 30 & 50 & 5 & 10 & 20 & 30 & 50 \\
\midrule
\multicolumn{11}{l}{\emph{Reasoning retrieval (informal $+$ formal query)}} \\
INF-X-Retriever                        & 21.2 & 28.2 & 33.9 & 40.0 & 42.6 & 15.9 & 18.8 & 21.7 & 24.6 & 24.6 \\
\quad $+$ query rewriter               & 13.2 & 17.4 & 25.2 & 28.4 & 38.3 & 10.1 & 10.1 & 17.4 & 17.4 & 24.6 \\
ReasonIR                               & 23.9 & 26.9 & 33.3 & 35.6 & 43.8 & 17.4 & 18.8 & 20.3 & 21.7 & 30.4 \\
DIVER (full pipeline)                  & 27.7 & 38.0 & 46.6 & 52.8 & 55.0 & 18.8 & 24.6 & 29.0 & 33.3 & 37.7 \\
\quad retriever only                   & 21.6 & 24.7 & 40.6 & 45.2 & 48.5 & 15.9 & 15.9 & 26.1 & 31.9 & 33.3 \\
\quad retriever $+$ query expansion          & 22.2 & 27.6 & 40.8 & 43.0 & 47.9 & 15.9 & 17.4 & 26.1 & 27.5 & 31.9 \\
\quad retriever $+$ GroupRank         & 29.7 & 38.5 & 46.2 & 52.8 & 55.0 & 18.8 & 24.6 & 27.5 & 33.3 & 37.7 \\
\textbf{LeanSearch~v2 (reasoning)}     & \textbf{36.7} & \textbf{46.1} & \textbf{50.8} & \textbf{56.4} & \textbf{57.1} & \textbf{27.5} & \textbf{30.4} & \textbf{34.8} & \textbf{43.5} & \textbf{43.5} \\
\midrule
\multicolumn{11}{l}{\emph{Premise selection (proof-state query)}} \\
ReProver                               & 3.0 & 5.5 & 7.6 & 8.8 & 10.3 & 0.0 & 1.4 & 2.9 & 4.3 & 4.3 \\
LeanPremise                            & 4.1 & 4.8 & 8.0 & 9.9 & 11.5 & 1.4 & 1.4 & 4.3 & 4.3 & 5.8 \\
LeanStateSearch                        & 6.5 & 9.3 & 12.5 & 16.0 & 20.0 & 2.9 & 2.9 & 7.2 & 10.1 & 13.0 \\
\bottomrule
\end{tabular}
\end{table}

LeanSearch~v2 (reasoning) leads every reported cell of
Table~\ref{tab:premise-main}: against the strongest reasoning
baseline (the DIVER full pipeline), v2 improves group-recall@$10$
by $+\,8.1$ points ($46.1$ vs.\ $38.0$) and Covered@$10$ by $+\,5.8$
points ($30.4$ vs.\ $24.6$). This margin reflects two design choices: reasoning mode plans at the granularity of an overall proof strategy rather than the next tactic step, yielding sub-queries that better match retrieval intent; and the judge--reviser loop re-routes the sketch when the corpus cannot support a planned premise, filtering out routes that are valid in principle but uncoverable in practice. Within the reasoning-retrieval block,
the INF-X query rewriter underperforms the bare INF-X retriever
(group-recall@$10$ of $17.4$ vs.\ $28.2$): the rewriter is trained to
distill verbose, reasoning-heavy queries into short search phrases,
but our queries are already a concise formal statement plus informal
description, so the distillation has little useful work to do and
tends to introduce paraphrasing drift. We also observe that within the reasoning-retrieval block the DIVER full pipeline sits closest to our reasoning-mode result---an outcome consistent with its own multi-stage design (query expansion, retrieval, listwise reranking) investing substantial inference-time LLM capacity, so the comparison runs in a regime where both sides commit comparable inference-time resources. The premise-selection block
recovers proportionally fewer premise groups across all $k$, which we read
as a scope-mismatch effect rather than a system-quality gap: the
premise-selection systems focus on finding premises that advance a
local proof state, and the set-valued target of global
premise retrieval falls outside that intended scope. We include them
for completeness as the dominant Lean-side retrieval line.
\FloatBarrier

\subsection{Prove: downstream proof-generation evidence}
\label{sec:exp-prove}

The Prove task tests whether the retrieval-quality ordering observed in \S\ref{sec:exp-premise} carries over to end-to-end proof generation. We adopt
a fixed proof-generation workflow --- the \emph{simple reflection loop} ---
in which a single prover LLM attempts the goal, and on each compile
failure the workflow routes the verifier trace through a
retrieve-and-reflect step before the next attempt. Across all
configurations the prover and reflection LLM are Claude Sonnet~4.5,
running with $8$ reflection rounds; the
criterion for counting a proof as solved is detailed in
Appendix~\ref{app:prove-filter}. We evaluate on FATE-H~\citep{fate}
($100$ algebra problems sampled across sub-fields) and on
\emph{MathlibMPR-Prop} ($50$ problems), the subset of MathlibMPR
consisting of theorems whose formal statement admits the uniform
tactic-mode format \texttt{:= by sorry} (Appendix~\ref{app:benchmark-curation}),
and which spans the same range of mathematical subjects as the
parent benchmark.

We compare six configurations: \emph{(i)} \textbf{no retrieval}
(prover sees only the goal); \emph{(ii)} \textbf{LeanFinder};
\emph{(iii)} \textbf{LeanSearch~v2 standard mode}; \emph{(iv)}
\textbf{LeanStateSearch}; \emph{(v)} \textbf{INF-X-Retriever}; and
\emph{(vi)} \textbf{LeanSearch~v2 reasoning mode}. Each retriever is
queried in the input format that matches its published contract: the
Lean-side semantic search engines (LeanFinder, LeanSearch~v2 standard
mode) accept a natural-language reformulation of the current goal;
LeanStateSearch accepts the formal proof state taken from the Lean~4
REPL at the first error in the prover's current attempt;
INF-X-Retriever and LeanSearch~v2 reasoning mode accept the informal
description $+$ formal statement of the theorem. The semantic search
engines fire only during reflection, since their queries depend on
the prover's current attempt; the retrievers that condition on the
theorem statement fire both at the initial attempt and during
reflection. Full input handling is documented in
Appendix~\ref{app:prove-integration}.

\begin{table}[!t]
\centering
\caption{\textbf{Prove task: end-to-end proof success under a fixed
simple reflection loop, varying only the retriever.} \emph{\#solved}
counts problems for which the workflow produced a Lean-verified,
sorry-free proof within $8$ reflection rounds.
FATE-H~\citep{fate} contains $100$ problems; MathlibMPR-Prop is
the $50$-problem subset of MathlibMPR consisting of theorems whose
formal statement admits the uniform tactic-mode \texttt{:= by sorry}
format (Appendix~\ref{app:benchmark-curation}). Best per column in
\textbf{bold}.}
\label{tab:prove-main}
\small
\setlength{\tabcolsep}{12pt}
\begin{tabular}{lcc}
\toprule
Retriever & FATE-H ($n{=}100$) & MathlibMPR-Prop ($n{=}50$) \\
\midrule
no retrieval                            & 4 (4.0\%)            & 2 (4.0\%)           \\
LeanFinder                              & 12 (12.0\%)          & 5 (10.0\%)          \\
LeanStateSearch                         & 7 (7.0\%)            & 3 (6.0\%)           \\
INF-X-Retriever                         & 16 (16.0\%)          & 5 (10.0\%)          \\
LeanSearch~v2 (standard mode)           & 14 (14.0\%)          & 5 (10.0\%)          \\
\textbf{LeanSearch~v2 (reasoning mode)} & \textbf{20 (20.0\%)} & \textbf{7 (14.0\%)} \\
\bottomrule
\end{tabular}
\end{table}

LeanSearch~v2 reasoning mode (Table~\ref{tab:prove-main}, last row)
attains the highest success rate on both datasets ($20.0\%$ on FATE-H,
$14.0\%$ on MathlibMPR-Prop), and the relative ordering among the four
retrievers that condition on the theorem statement matches the
MathlibMPR group-recall ranking from Table~\ref{tab:premise-main}.
The two Lean-side semantic search engines, queried only at reflection
time, sit between the no-retrieval baseline and the retrievers that
see the full theorem statement up front. This concludes our
evaluation: retrieval quality at the candidate-set level propagates to
end-to-end proof success rather than being attenuated by the prover
and reflection loop, with the relative downstream ordering matching
the upstream retrieval ordering across datasets.
\FloatBarrier

\section{Limitations}

First, we evaluate exclusively on Lean~4/Mathlib. The methodology is language-agnostic in design, but empirical transfer to other provers (Coq, Isabelle) with different library conventions remains to be demonstrated.
Second, our downstream experiment isolates retrieval quality by using a minimal reflection loop workflow designed to prove a single theorem at a time.
How the system behaves inside highly capable agent systems~\citep{seedprover,aristotle,archon} on complex tasks such as repository-level formalization, where retrieval interacts with planning and tactic generation, is unexplored.
Third, we focus on retrieval for proof construction. Other retrieval-intensive Lean tasks, such as locating canonical definitions or navigating the type-class hierarchy during statement construction, may require different query formulations and evaluation criteria and are not studied here.

\section{Conclusion}

We formulated \emph{global premise retrieval} and presented LeanSearch~v2, a staged system whose standard mode achieves state-of-the-art Mathlib search without domain-specific fine-tuning and whose reasoning mode targets set-valued premise retrieval through iterative sketch-retrieve-reflect cycles. Two expert-curated benchmarks, MathlibQR and MathlibMPR, provide the evaluation foundation; to our knowledge, this work is the first to bring reasoning-retrieval techniques to the Lean premise retrieval problem. Reasoning mode leads all baselines on MathlibMPR, and a controlled downstream experiment confirms that this advantage propagates to proof success. We hope the task formulation, benchmarks, and system presented here help establish retrieval as a distinct axis of progress for formal mathematics.

\section*{Acknowledgments}
This work is supported in part by the National Key R\&D Program of China grant 2024YFA1014000, the Fundamental and Interdisciplinary Disciplines Breakthrough Plan of the Ministry of Education of China (JYB2025XDXM113), and the New Cornerstone Investigator Program.

{\sloppy
\bibliographystyle{plainnat}
\bibliography{references}
\par}


\appendix


\newpage

\section{Benchmark Curation}
\label{app:benchmark}

In this section, we present the curation methodology, key characteristics, and illustrative examples of our benchmark to provide readers with a clearer understanding of both the benchmark and the experimental results. We introduce two benchmarks: \textbf{MathlibQR} for the classical formal theorem searching task, and \textbf{MathlibMPR} for global premise retrieval. Detailed descriptions are provided in the appendix: MathlibQR is discussed in Appendix~\ref{app:mathlibqr}, and MathlibMPR in Appendix~\ref{app:benchmark-curation}.

\subsection{Searching Mathlib theorems: MathlibQR}
\label{app:mathlibqr}

\textbf{MathlibQR} is designed for classical theorem retrieval tasks: given a user query, the goal is to select the most relevant formal theorem (document).

\subsubsection{Curation and characteristics}

The benchmark comprises 200 documents and 946 queries, where multiple queries may correspond to a single document. This design reflects the diversity of human or AI generated search queries. Each document can be associated with up to six distinct query styles, namely Lean-centric, LaTeX expressions, pure natural language, slogan, nickname, and special instances, with at most one query per style. This variety specifically captures the heterogeneity of Lean users with different levels of familiarity with Lean declarations and abstraction, as well as mathematicians of varying maturity, who may express queries using natural language, LaTeX formulas, slogans, or nicknames.

The last category, \emph{special instances}, addresses a common but previously underexamined challenge: users often have a specific application scenario in mind without knowing the exact formulation of the corresponding statement in Mathlib. For instance, a user may wish to apply a theorem about convex sets in $\mathbb{R}^n$ without being aware of its generalization in Mathlib. In this case, the actual condition replacing $\mathbb{R}^n$ involves \verb`[Semiring k] [PartialOrder k] [AddCommMonoid E] [SMul k E]`, which requires expert formalization knowledge to anticipate. A well-designed formal theorem search engine should return relevant results even in such demanding cases. Additional examples of different query styles are provided in Appendix~\ref{app:mathlibqr-example}.

In terms of coverage, the benchmark spans a wide range of mathematical domains—including algebra, analysis, logic, geometry, probability, number theory, topology, and category theory—and covers all major Mathlib declaration types, such as theorems, definitions, and instances. Each document is further classified into two difficulty levels, easy and hard, based on two dimensions: mathematical obscurity and technical complexity.

To construct such a broadly representative benchmark, we first asked formalization experts to select 8 documents from each of 25 top-level Mathlib folders (out of 31 total, excluding \verb`Control`, \verb`Deprecated`, \verb`Lean`, \verb`Tactic`, \verb`Testing`, and \verb`Util`), ensuring balanced coverage across declaration types and difficulty levels. After finalizing the 200 documents, the experts produced up to six query styles per document, wherever a given style was applicable.

In total, we collected 946 queries (detailed statistics are provided in Table~\ref{tab:mathlibqr-stats}), based on Mathlib version v4.29.1, the latest release at the time of construction. As Mathlib is continuously evolving, the corpora of LeanFinder, LeanExplore and this work are built on different versions. Consequently, a subset of 171 documents with 810 queries falls within the common corpus shared by all search tools compared in our experiments (Section~\ref{sec:exp-search}).

\begin{table}[ht]
\centering
\caption{Benchmark statistics of MathlibQR across the 200 entries.}
\label{tab:mathlibqr-stats}

\begin{subtable}[t]{\textwidth}
\centering
\caption{By query type}
\label{tab:query-stats-type}
\begin{tabular}{ccccccc}
\toprule
&Lean & LaTeX & Natural & Slogan & Nickname & Special Case \\
\midrule
Query Number & 199 & 200 & 199 & 197 & 128 & 23 \\
\bottomrule
\end{tabular}
\end{subtable}

\vspace{1em}

\begin{subtable}[t]{\textwidth}
\centering
\caption{By difficulty}
\label{tab:query-stats-diff}
\begin{tabular}{ccc}
\toprule
&Easy & Hard \\
\midrule
Document Number&101 & 99 \\
\bottomrule
\end{tabular}
\end{subtable}

\end{table}

\subsubsection{Examples}
\label{app:mathlibqr-example}

These examples illustrate how a mathematical result can be expressed through diverse query styles, reflecting the varied ways users—from formalization experts to mathematicians—might search for the same theorem. Each document in the benchmark is paired with four to six such styles, capturing the heterogeneity of search intents in practice.

\paragraph{Example 1. \texttt{Finset.all\_card\_le\_biUnion\_card\_iff\_existsInjective$^\prime$}.} Hall's marriage theorem, a classic result in combinatorics, states that a perfect matching exists between $n$ men and $n$ women if and only if every $k$-subset of men collectively likes at least $k$ women. Combinatorial theorems of this kind frequently arise in application-oriented contexts, motivating us to include a concrete special case as a query.

\begin{table}[ht]
\centering
\caption*{1. \texttt{Finset.all\_card\_le\_biUnion\_card\_iff\_existsInjective$^\prime$}}
\label{tab:example-cb6}
\begin{tabular}{p{3cm} p{10cm}}
\toprule
Lean & \texttt{(\(\forall\) s : Finset \(\iota\), \#s \(\le\) \#(s.biUnion t)) \(\iff\) \(\exists\) f : \(\iota\) \(\to\) \(\alpha\), Injective f \(\wedge\) \(\forall\) x, f x \(\in\) t x} \\
LaTeX & $\forall S, |S| \le \left|\bigcup_{i \in S} t(i)\right| \iff \exists f$ injective, $f(i) \in t(i)$ \\
Natural & A finite system has a system of distinct representatives iff every subfamily covers enough elements \\
Slogan & Hall's condition characterizes existence of an SDR \\
Nickname & Hall's marriage theorem \\
Special case & $n$ men and $n$ women have a perfect matching iff every $k$-subset of men collectively likes at least $k$ women \\
\bottomrule
\end{tabular}
\end{table}

\paragraph{Example 2. \texttt{exists\_continuous\_zero\_one\_of\_isClosed}.} Urysohn's lemma states that for any two disjoint closed sets $S$ and $T$ in a normal topological space $X$, there exists a continuous function $f : X \to \mathbb{R}$ such that $f|_S = 0$, $f|_T = 1$, and $0 \leq f(x) \leq 1$ for all $x \in X$. This theorem is widely recognized both by its name and its slogan. We also include a special case applying it to the real numbers—a query style for which recovering the general normal-space formulation is nontrivial.

\begin{table}[ht]
\centering
\caption*{2. \texttt{exists\_continuous\_zero\_one\_of\_isClosed}}
\label{tab:example-t7}
\begin{tabular}{>{\raggedright\arraybackslash}p{3cm} >{\raggedright\arraybackslash}p{10cm}}
\toprule
Lean & \texttt{for disjoint closed s t in a NormalSpace, exists continuous f with f = 0 on s, f = 1 on t} \\
LaTeX & $\exists f \in C(X, [0,1]),\ f|_S = 0,\ f|_T = 1$ for $S, T$ disjoint closed in normal $X$ \\
Natural & In a normal space, two disjoint closed sets can be separated by a continuous function valued 0 on one and 1 on the other \\
Slogan & Disjoint closed sets in a normal space are separated by a continuous function \\
Nickname & Urysohn's lemma \\
Special case & In $\mathbb{R}$, $[0,1]$ and $[2,3]$ are separated by a continuous ramp \\
\bottomrule
\end{tabular}
\end{table}

\paragraph{Example 3. \texttt{AlexandrovDiscrete}.} An Alexandrov-discrete space (or Alexandrov topology) is a topological space in which the intersection of every family of open sets is open—a condition strictly weaker than discreteness. Since this name is less familiar than the property itself, the definition is more naturally searched for via descriptive styles (Lean, LaTeX, natural language) than by nickname.

\begin{table}[ht]
\centering
\caption*{3. \texttt{AlexandrovDiscrete}}
\label{tab:example-t3}
\begin{tabular}{>{\raggedright\arraybackslash}p{3cm} >{\raggedright\arraybackslash}p{10cm}}
\toprule
Lean & \texttt{a TopologicalSpace where \(\bigcap\!_{0}\) S is open for any family S of opens} \\
LaTeX & $\bigcap_{i \in I} U_i$ open for arbitrary index set $I$ \\
Natural & A topological space where intersections of arbitrary open families remain open \\
Slogan & Arbitrary intersections of opens stay open \\
Nickname & Alexandrov-discrete space \\
Special case & --- \\
\bottomrule
\end{tabular}
\end{table}

\paragraph{Example 4. \texttt{Condensed.instAB4CondensedMod}.} Some statements are stored as instances rather than theorems. The slogan "CondensedMod satisfies AB4" captures one such example. We also include the Lean query \texttt{instance : AB4 (CondensedMod R)}, which is the most direct way for formalization experts to search for this result.

\begin{table}[ht]
\centering
\caption*{4. \texttt{Condensed.instAB4CondensedMod}}
\label{tab:example-cd6}
\begin{tabular}{p{3cm} p{10cm}}
\toprule
Lean & \texttt{instance : AB4 (CondensedMod R)} \\
LaTeX & $\mathrm{CondensedMod}\,R$ satisfies Grothendieck axiom AB4 \\
Natural & Condensed modules satisfy Grothendieck AB4 (infinite direct sums are exact) \\
Slogan & CondensedMod satisfies AB4 \\
Nickname & --- \\
Special case & --- \\
\bottomrule
\end{tabular}
\end{table}

\subsection{Global premise retrieval: MathlibMPR}
\label{app:benchmark-curation}

\textbf{MathlibMPR} is a benchmark for global premise retrieval: given a mathematical problem, the task is to retrieve the theorems representing the key mathematical steps that, when combined, yield a solution. The name MathlibMPR derives from its construction method—the benchmark is built from merged pull requests in Mathlib.

\subsubsection{Curation and characteristics}

MathlibMPR contains 69 mathematical problems extracted from merged Mathlib pull requests. Each problem is equipped with a natural language statement, a formal statement, and a list of premise groups. Each premise group consists of several lemmas that are mathematically equivalent but differ in their formalizations, and is tagged as either \emph{original} or \emph{alternative}. The original tag indicates lemmas that appear as vital steps in the proof merged into Mathlib, while the alternative tag indicates lemmas that play an essential role in a different, equally viable proof.

During curation, formalization experts selected merged PRs that postdate the corpora of all search engines compared in our experiments (Section~\ref{sec:exp-premise}). For each selected PR, experts identified a main theorem, summarized the key mathematical steps, and annotated the corresponding Mathlib declarations. When possible, they also constructed alternative proofs viable within Mathlib and annotated the lemmas constituting the key steps of those proofs.

All selected PRs were merged more than six months after the knowledge cutoff of the prover model used in our experiments (Claude Sonnet 4.5, Section~\ref{sec:exp-prove}), ensuring no data contamination.

To ensure compatibility with the prover pipeline, we further filtered the problems down to 50 formal statements with a uniform format (each ending in \verb`:= by sorry`), allowing all premise selection tools to function correctly.

\subsubsection{Examples}

In this section, we highlight several key properties of global premise retrieval: navigating typeclass inheritance chains, proposing intermediate reduction steps, accommodating multiple viable proof strategies, and using abstract machinery. Some of these properties are shared with general reasoning retrieval tasks, while others are distinctive to formal mathematics; we will provide a detailed comparison later in Appendix~\ref{app:bright}.

\paragraph{Example 1 (Intermediate reduction steps).} PR 30112 proves Kallenberg's "isolation of randomness" lemma: given a standard Borel space $Y$ and a Markov kernel $\kappa : X \rightsquigarrow Y$, there exists a jointly measurable map $f : X \to I \to Y$ such that for every $a : X$, the pushforward of the uniform measure on the unit interval $I$ via $f(a, \cdot)$ equals $\kappa(a, \cdot)$.

The proof requires first reducing to the case $Y = I$, using the fact that every Markov kernel into a standard Borel space can be represented as the pushforward of the uniform distribution on $[0,1]$ under a family of jointly measurable functions. Without this reduction, the technical lemmas for constructing the inverse CDF are inapplicable. This reduction step is highly nontrivial: it does not emerge naturally from the original statement and demands substantial mathematical experience and, often, multiple reasoning attempts to discover.

\paragraph{Example 2 (Multiple proof strategies).} PR 36490 proves the following statement in category theory: given an adjunction $F \dashv G : \mathcal{C} \rightleftarrows \mathcal{D}$ with Grothendieck topologies $J$ on $\mathcal{C}$ and $K$ on $\mathcal{D}$, the left adjoint $F$ is cocontinuous from $J$ to $K$ (every $K$-cover of $F U$ pulls back to a $J$-cover of $U$) if and only if $G$ preserves covers from $K$ to $J$ (every $K$-cover of $X$ is sent to a $J$-cover of $G X$).

The proof merged in the PR proceeds by elementary manipulations of sieves using the unit and counit of the adjunction. A conceptually cleaner alternative is available: $F$ is cocontinuous if and only if its pullback $G^*$ preserves sheaves, which holds if and only if $G$ preserves covers. This route is more elegant but requires substantial sheafification machinery beyond what is currently available in Mathlib, making it impractical to execute within a reasonable proof length. Consequently, the elementary approach was adopted.

\paragraph{Example 3 (Abstract machinery).} PR 31754 proves that every non-empty light profinite space is an injective object in the category of profinite spaces: for $S$ light profinite and non-empty, any commutative square
\[
\begin{tikzcd}
X \arrow[r, hook] \arrow[d, "g"'] & Y \arrow[ld, dashed] \\
S
\end{tikzcd}
\]
with $X \hookrightarrow Y$ a monomorphism in the category of profinite spaces admits a lift $Y \to S$ extending $g$.

The proof proceeds in two stages. First, one establishes the result for finite spaces $S$ and maps between finite spaces. Then, writing any light profinite space $S$ as a cofiltered sequential limit of finite spaces, one applies the abstract machinery of ``passing to limits'' to transfer the result to the general case. Both preliminary steps are specifically crafted to feed into this abstract argument: without recognizing that the problem calls for passage to a cofiltered limit, the individual lemmas serve no purpose. Solving this problem thus demands the ability to abstract from the concrete statement to the framework of cofiltered limits—a step that depends crucially on the capacity to recognize the right abstract structure underlying a concrete problem.

\subsubsection{Comparison with reasoning retrieval tasks}
\label{app:bright}

The previous section highlighted several challenges unique to global premise retrieval: proposing intermediate reduction steps, choosing among multiple proof strategies, and using abstract machinery. In this section, we compare these characteristics with those of BRIGHT~\citep{bright}, a traditional reasoning retrieval benchmark. The following two examples are drawn from its appendix.

\paragraph{Example 1 (Table 27 in~\citet{bright}).} This example is drawn from a LeetCode problem:

\textbf{Query:}
\begin{quote}
Given `n' non-negative integers representing an elevation map where the width of each bar is `1', compute how much water it can trap after raining.

\textbf{Example 1:}\\
\textbf{Input:} \texttt{height = [0, 1, 0, 2, 1, 0, 1, 3, 2, 1, 2, 1]} \\
\textbf{Output:} 6 \\
\textbf{Explanation:} The above elevation map (black section) is represented by array \texttt{[0, 1, 0, 2, 1, 0, 1, 3, 2, 1, 2, 1]}. In this case, 6 units of rain water (blue section) are being trapped.

\textbf{Example 2:}\\
\textbf{Input:} \texttt{height = [4, 2, 0, 3, 2, 5]} \\
\textbf{Output:} 9

\textbf{Constraints:}
\begin{itemize}
  \item \texttt{n == height.length}
  \item \texttt{1 <= n <= 2 * 10\textsuperscript{4}}
  \item \texttt{0 <= height[i] <= 10\textsuperscript{5}}
\end{itemize}
\end{quote}

\textbf{Chain-of-thought reasoning to find documents:}
\begin{quote}
This problem can be solved using a two-pointer approach, and uses ideas from dynamic programming to keep track of the maximum height to the left and right of each bar. We can find other example code that also use these techniques.
\end{quote}

\textbf{Analysis of Example 1.} This retrieval problem also requires reasoning before retrieval: one must first recall possible solution strategies (e.g., two-pointer techniques, dynamic programming) before searching for relevant documents. However, it lacks the distinctive challenges that appear in global premise retrieval—namely, the need to abstract from a concrete statement to a general framework, or to choose among multiple viable proof strategies based on what machinery is available. The reasoning here relies more on association and recognition than on structural abstraction, making it less demanding in terms of reasoning depth than formal theorem retrieval.

\paragraph{Example 2 (Table 29 in~\citet{bright}).} This example is drawn from 2015 AMC 10B Problem 15:

\textbf{Query}:
\begin{quote}
The town of Hamlet has 3 people for each horse, 4 sheep for each cow, and 3 ducks for each person. Which of the following could not possibly be the total number of people, horses, sheep, cows, and ducks in Hamlet?
(A) 41 (B) 47 (C) 59 (D) 61 (E) 66
\end{quote}

\textbf{Chain-of-thought reasoning to find documents:}
\begin{quote}
We can use the Chicken McNugget Theorem to solve this problem. We can find other solutions that also apply this theorem.
\end{quote}

\textbf{Analysis of Example 2.} 
This retrieval problem hinges on a single, well-known theorem: once the solver recognizes the problem as an application of the Chicken McNugget Theorem, the retrieval target is essentially fixed. There is no need to find the reduction steps, evaluate multiple viable proof strategies, or lift a concrete statement to an abstract framework. The reasoning required in this task is considerably less demanding than research-level formal premise retrieval in Mathlib.

\section{Experiment Setup}

\subsection{Reasoning-mode aggregation across sub-queries}
\label{app:reasoning-aggregation}

We rank the final outputs of reasoning mode as follows. A reasoning-mode
sketch decomposes the target theorem into multiple sub-queries; each
sub-query goes through standard mode and is post-processed by the
document filter, yielding a per-sub-query ranked list of accepted
Mathlib declarations. The same declaration may appear in several
sub-queries' lists, and across sub-queries the raw retrieval scores
are not directly comparable, since each sub-query is its own retrieval
distribution under a different instruction. We therefore discard the
raw scores and aggregate by rank position only.

Concretely, let $L_1, \dots, L_m$ be the per-sub-query filtered lists
for an accepted sketch (each $L_j$ is a length-$|L_j|$ list of
declaration ids). For each declaration $d$ that appears at $0$-indexed
rank $i_j(d)$ in $L_j$ we assign a positional contribution
$1 / \log_2(i_j(d) + 2)$ from list $j$, and define $d$'s aggregate
score as
\[
  s(d) \;=\; \sum_{j \,:\, d \in L_j} \frac{1}{\log_2\!\bigl(i_j(d) + 2\bigr)}.
\]
The $j$-th list contributes $1.00$ for its rank-$1$ entry, $0.63$ for
rank $2$, $0.50$ for rank $3$, and so on, with the same declaration's
contributions summed across the lists in which it appears. Reasoning
mode's final ranking is the descending sort of $\{s(d)\}$ over the
union of accepted declarations, truncated to top-$k$.

We adopted this rank-only rule after an initial design that aggregated
by taking each declaration's maximum raw cosine or rerank score across
sub-queries. The score-based design was empirically poor at low $k$:
the embedder's cosine similarities and the reranker's $P(\texttt{yes})$
outputs are conditioned on per-sub-query instructions and queries, so
scores on different sub-queries occupy incomparable ranges. Replacing
raw scores with a position-only discount removes the cross-sub-query
calibration problem entirely and gives consistent improvements at
small $k$ without sacrificing performance at $k\!\ge\!100$.

\subsection{LLM-as-judge protocol for the Search task}
\label{app:llm-judge}

The LLM-as-judge column in Table~\ref{tab:search-main} follows the
protocol introduced for Mathlib search by \citet{leanexplore}, with
two modifications below to accommodate a four-system comparison and
to mitigate position bias. We briefly explain how the protocol works
and what self-checks we ran.

Each retrieved declaration is presented to the judge as a six-field
block: fully-qualified Mathlib name, declaration kind, human-readable
informal name, formal signature (truncated to $800$ characters),
value or proof body (truncated to $400$ characters), and informalized
natural-language statement (truncated to $600$ characters). All four
systems' results are hydrated from a single shared metadata source
used by standard mode, so the form of the judge's reading material
is identical across systems. For each of the $810$ active queries we
sample three distinct permutations (without replacement, out of $24$
possible) of the four systems via a deterministic per-query seed,
and in every prompt the systems are presented anonymously as
\emph{System A} / \emph{B} / \emph{C} / \emph{D} so that the judge
cannot exploit system identity; the per-query permutations across
rounds expose each system to multiple label positions. The judge is
Claude Sonnet~4.5, called at \texttt{temperature}$=0$,
\texttt{max\_tokens}$=600$, and prompted to produce a strict total
ordering $1$--$4$ over the four top-$5$ result sets, output as a
JSON object whose \texttt{ranking} field is a permutation of
\verb|["A","B","C","D"]|. Ties are forbidden by prompt; the parser
observed no malformed responses across $2430$ judgments per system
after stripping JSON code fences. For each (slice, system) cell we
report mean rank (lower is better; range $1$--$4$) over
$n_{\text{slice}} \times 3$ judgments, which on the overall slice is
$810 \times 3 = 2430$ judgments per system.

Two self-checks support reading the resulting numbers as stable. The
first is cross-round agreement: per-round mean ranks vary by at most
a few hundredths of a rank across the three rounds, and the mean
pairwise Spearman correlation between the round-$i$ and round-$j$
rank vectors is consistently high. The second is a position-bias
check: the judge does exhibit a primacy bias --- every system
receives a better mean rank when assigned to label position \emph{A}
than to \emph{D}, with a swing of approximately $0.42$--$0.58$ ranks
(Table~\ref{tab:llm-judge-position-bias}) --- but the permutation
procedure assigns each system to each of the four label positions
approximately equally often ($574$--$640$ assignments per cell on
$2430$ judgments; uniform expectation is $607.5$), so the bias
averages out at the system level and the reported overall mean ranks
are debiased.

\begin{table}[ht]
\centering
\caption{Mean rank conditioned on the system's randomized label position. Values are means over the queries assigned to that position by the per-query seed.}
\label{tab:llm-judge-position-bias}
\small
\setlength{\tabcolsep}{6pt}
\begin{tabular}{lcccc}
\toprule
System & Position A & Position B & Position C & Position D \\
\midrule
LeanSearch~v2 (rerank)         & 1.290 & 1.642 & 1.718 & 1.865 \\
LeanSearch~v2 (retriever only) & 1.854 & 2.156 & 2.327 & 2.370 \\
LeanFinder                     & 2.522 & 2.877 & 3.091 & 2.985 \\
LeanExplore                    & 3.011 & 3.379 & 3.479 & 3.429 \\
\bottomrule
\end{tabular}
\end{table}

\subsection{Global premise retrieval: per-system input and output handling}
\label{app:retriever-io}

The eight reasoning-retrieval rows and three premise-selection rows
in Table~\ref{tab:premise-main} differ in two procedural respects:
what they receive as the per-query input, and how their output is
post-processed before scoring. Within each family all systems
produce a top-$100$ list of Mathlib declaration ids, and Mathlib
declaration names are unique once normalised, so output handling is
not a per-system distinction but a uniform doc-id match. Input,
however, splits into three families --- our reasoning-mode pipeline,
the reasoning retrievers, and the premise-selection systems --- and
we describe each family in turn. Table~\ref{tab:premise-system-io}
summarises the resulting per-row pipelines.

\paragraph{LeanSearch~v2 reasoning mode.}
We treat reasoning mode (\S\ref{subsec:reasoning}) as a black box
here: input is the raw \texttt{query} field --- the concatenation of
the informal description and the formal Lean statement --- and output
is a ranked list of Mathlib declaration ids of variable length. Mean
output length on \textbf{MathlibMPR} is $30.6$ docs per query
(P25 $=21$, median $=26$, P75 $=38$, max $=90$); we score the system
at every $k \le 100$ but its $k\!\ge\!50$ cells are nearly identical
to its $k\!=\!30$ cells because the system is output-constrained at
large $k$ and does not pad to a fixed length.

\paragraph{Reasoning retrievers.}
The reasoning-retrieval rows take the same raw \texttt{query} as
input and feed it directly to the corresponding encoder, except for
two systems that wrap the query before retrieval. INF-X-Retriever
in its rewriter regime runs \texttt{infly/inf-query-aligner} to
produce a single short search query before passing it to the
\texttt{infly/inf-retriever-v1-pro} encoder. DIVER in its
qexpand regime runs \texttt{DeepSeek-R1-Distill-Qwen-14B} as a
query-expansion reasoner whose output is concatenated with the raw
query and fed to the \texttt{Diver-Retriever-4B} encoder. The other
INF-X, ReasonIR, and DIVER configurations feed the raw query
directly. Output is the encoder's (or rerank-augmented encoder's)
top-$100$ over the same Mathlib corpus we use for standard mode.

\paragraph{Premise selection.}
ReProver, the \texttt{lean-premises} encoder, and
LeanStateSearch all consume a \emph{proof state} (goal $+$ hypothesis
context) rather than an informal description, matching their
published query contracts. We extract the proof state by appending
\texttt{:= sorry} to each query's formal statement and re-running the
resulting declaration through the Lean~$4$ REPL via
LeanInteract~\citep{leaninteract}: the REPL emits a \texttt{Sorry}
record whose \texttt{goal} field is the target proof state. Coverage on \textbf{MathlibMPR} is
$69/69$ queries successfully extracted. Each system retrieves the
top-$100$ from its own preferred premise corpus.

\begin{table}[ht]
\centering
\caption{\textbf{Per-system pipelines on MathlibMPR.} \emph{Input}: I/F $=$ informal $+$ formal statement; PS $=$ extracted proof state. \emph{Corpus}: Mathlib $=$ our informalized Mathlib snapshot; LD $=$ LeanDojo Benchmark premises; LP $=$ the premise corpus released with LeanPremise~\citep{leanhammer}; LSS $=$ premise-search.com server-side corpus.}
\label{tab:premise-system-io}
\small
\setlength{\tabcolsep}{4pt}
\begin{tabular}{llll}
\toprule
System (regime) & Input & Pipeline & Corpus \\
\midrule
INF-X retriever-only          & I/F & retrieve top-$100$                          & Mathlib \\
INF-X $+$ rewriter            & I/F & query-aligner $\to$ retrieve top-$100$       & Mathlib \\
ReasonIR                      & I/F & retrieve top-$100$                          & Mathlib \\
DIVER retriever               & I/F & retrieve top-$100$                          & Mathlib \\
DIVER $+$ qexpand             & I/F & R1-Distill-14B expand $\to$ retrieve top-$100$ & Mathlib \\
DIVER $+$ rerank              & I/F & retrieve top-$100$ $\to$ GroupRank-32B listwise rerank & Mathlib \\
DIVER full pipeline           & I/F & qexpand $\to$ retrieve $\to$ rerank, score-blended & Mathlib \\
\textbf{LeanSearch~v2 reasoning} & I/F & sketch (Sonnet) $\to$ leansearch retriever\,$\to$ & Mathlib \\
& & filter (Kimi K2 Instruct) $\to$ judge (Sonnet) $\to$ &  \\
& & rank-discount aggregate &  \\
\midrule
ReProver                      & PS & byt5-small encoder $\to$ retrieve top-$100$  & LD \\
LeanPremise                   & PS & all-distilroberta-v$1$ ft encoder $\to$ retrieve top-$100$ & LP \\
LeanStateSearch               & PS & HTTP API $\to$ top-$100$ & LSS \\
\bottomrule
\end{tabular}
\end{table}

\subsection{Prove: per-system input handling within the simple reflection loop}
\label{app:prove-integration}
\label{app:prove-filter}

Every row of Table~\ref{tab:prove-main} runs the same simple
reflection loop: a prover LLM emits a candidate Lean proof, the Lean
REPL verifies it, and on a compile failure a reflection step
regenerates the proof with the REPL feedback (and, optionally,
retrieval hints) appended. Per-row differences live in two slots:
what the retriever sees as its query, and at which step of the loop
the retriever fires. We describe each row below, then summarise the
integration in Table~\ref{tab:prove-system-io}. All rows share the
same loop hyperparameters --- $8$ reflection rounds,
\texttt{prover\_max\_retries}$=3$,
\texttt{verifier\_max\_retries}$=3$,
\texttt{verifier\_wait}$=30$ seconds --- and the prover LLM is
Claude Sonnet~4.5 in every row of Table~\ref{tab:prove-main};
per-row Kimi K2 Instruct-as-prover ablations are reported in
Appendix~\ref{app:prove-ablation}. A proof counts as solved iff it
compiles and contains no live \texttt{sorry} (block- and line-comments
are stripped before the check).

\paragraph{No retrieval.}
The prover sees only the informal description and the formal
statement on round one, and on each reflection round receives the
previous broken proof and the REPL error message. There is no
retriever in this configuration; the row anchors the contribution
that retrieval makes when added to the same loop.

\paragraph{LeanSearch~v2 (standard mode).}
On each reflection round, an LLM rewrites the prover's current
attempt into a natural-language query, queries standard mode, and
concatenates the top hits into the reflection prompt. The query
rewrite is necessary because standard mode is designed to consume
natural-language search phrases, whereas the prover's intermediate
state is a partial Lean proof together with compilation errors.

\paragraph{LeanFinder.}
Mechanically identical to the LeanSearch~v2 standard-mode row except
that the natural-language query is sent to the LeanFinder Hugging
Face Space via \texttt{gradio\_client}'s \texttt{/retrieve} endpoint,
and the HTML-table response is parsed into the canonical hit schema
before being inserted into the reflection prompt.

\paragraph{LeanStateSearch.}
On every round (including round one) the loop extracts the formal
proof state through the Lean REPL: it appends
\texttt{:= sorry} to the current attempt, runs it, and reads the
goal/hypothesis context from the resulting \texttt{Sorry} record.
That state string is sent to the LeanStateSearch HTTP API, and the
returned premise list is inserted into the next prover prompt. We
do not run a query-rewrite step here because LeanStateSearch is
designed and trained to consume formal proof states directly.

\paragraph{INF-X-Retriever.}
On every round (including round one) the loop sends a raw query
string to a local FastAPI service that wraps the INF-X-Retriever
encoder. On round one the query is the concatenation of the informal
description and the formal statement, without the rewriter provided
by INF-X-Retriever (see \S\ref{sec:exp-premise} for the rewriter's
effect on retrieval quality); on reflection rounds the query is the
concatenation of the current Lean code and the REPL error message. The
service returns doc ids only, which the loop hydrates against our
metadata pickle to fill in the formal signature and informalized
statement that the prover prompt expects.

\paragraph{LeanSearch~v2 (reasoning mode).}
On round one the loop runs reasoning mode (\S\ref{subsec:reasoning})
on the informal description $+$ formal statement, producing a sketch
and an aggregated premise set; the sketch text and the aggregated
premise list are both inserted into the prover prompt. On reflection
rounds the loop falls back to the standard-mode reflection path: an
LLM rewrites the prover's current attempt into a natural-language
query, queries standard mode, and feeds the hits back. We do not
re-invoke reasoning mode at every reflection round because the cost
would be prohibitive; the reasoning-mode output is computed once and
treated as a fixed sketch that the prover refines against.

\begin{table}[ht]
\centering
\caption{\textbf{Retriever integration across Prove-task rows.} \emph{Round 1}: input to retriever on the first prover call. \emph{Reflection}: input on subsequent rounds after a compile failure. ``--'' indicates no retrieval call at that step. \textit{informal/formal} $=$ the theorem's informal description $+$ formal statement; \textit{NL queries} $=$ an LLM-produced natural-language query distilled from the prover's current attempt; \textit{proof state} $=$ the formal proof state extracted from the Lean~4 REPL; \textit{code+err} $=$ the prover's current attempt concatenated with the REPL error message.}
\label{tab:prove-system-io}
\small
\setlength{\tabcolsep}{2pt}
\begin{tabular}{llll}
\toprule
Row & Round-$1$ input & Reflection input & Notes \\
\midrule
no retrieval                       & --                 & --                  &  \\
LeanSearch~v2 (standard mode)      & --                 & NL queries          & retriever fires only on reflection \\
LeanFinder                         & --                 & NL queries          & gradio\_client to HF Space \\
LeanStateSearch                    & proof state              & proof state               & proof state extracted at every round \\
INF-X-Retriever                    & informal $+$ formal & code $+$ err        & raw query, BRIGHT instruction \\
\textbf{LeanSearch~v2 (reasoning)} & informal $+$ formal & NL queries          & reasoning sketch frozen after round $1$ \\
\bottomrule
\end{tabular}
\end{table}

\section{More Experiment Results}

\subsection{Reasoning-mode loop iterations on MathlibMPR}
\label{app:reasoning-loop-stats}

We instrument reasoning mode on the \textbf{MathlibMPR}
benchmark to record, per query, which initial-or-revised sketch the
judge accepted (if any), how many revisions were required, and
whether the parallel branches agreed. Each query is run with
$\text{budget}\!=\!2$ branches in parallel and a per-branch revision
cap of $3$, so each branch issues an initial sketch and up to three
revised sketches before timing out. The two statistics we report are
the revision count of the accepted sketch and the joint outcome of
the two branches.

The judge accepts at least one branch's sketch for $46$ of the $69$
queries ($66.7\%$); the remaining $23$ queries ($33.3\%$) reach the
revision cap on every branch, in which case the per-sub-query filtered lists from both branches are pooled by the rank-discount rule of Appendix~\ref{app:reasoning-aggregation}.
Among the $46$ accepted queries, the winning branch's revision count
is distributed as in Table~\ref{tab:loop-revision-depth}: nearly
two-thirds of accepted queries are accepted on the initial sketch
with no revision, and no accepted query requires the third revision.

\begin{table}[ht]
\centering
\caption{\textbf{Revision count of the accepted sketch.} \emph{Initial} = sketch generator only; \emph{revised once / twice} = sketch reviser invoked one or two times before acceptance. Among the $23$ unaccepted queries every branch exhausts the cap (revised three times).}
\label{tab:loop-revision-depth}
\small
\setlength{\tabcolsep}{8pt}
\begin{tabular}{lcc}
\toprule
Revision count of accepted sketch & \# queries & \% of $69$ \\
\midrule
$0$ (initial sketch accepted)              & 29 & 42.0\% \\
$1$ (one revision before acceptance)       & 11 & 15.9\% \\
$2$ (two revisions before acceptance)      &  6 &  8.7\% \\
\midrule
\textit{Unaccepted (no branch accepted within $3$ revisions)} & 23 & 33.3\% \\
\bottomrule
\end{tabular}
\end{table}

At the branch level, individual branches independently terminate as
\texttt{good} (judge-accepted), \texttt{early\_stop} (the runtime
detected that a sibling's accepted output already satisfied the
budget and halted further work), or \texttt{fail} (revision cap
exhausted without acceptance). The joint outcomes across the two
branches are listed in Table~\ref{tab:loop-branch-joint}, and among
the $46$ accepted queries branch $0$ produced the winning sketch on
$31$ queries while branch $1$ produced it on $15$ --- a slight
imbalance consistent with branch $0$ being initialised first and so
more often reaching the judge before \texttt{early\_stop} fires on
its sibling.

\begin{table}[ht]
\centering
\caption{\textbf{Joint branch-status distribution.} Branch order is the launch order; \texttt{early\_stop} indicates the runtime cancelled the branch once a sibling produced an accepted sketch.}
\label{tab:loop-branch-joint}
\small
\setlength{\tabcolsep}{10pt}
\begin{tabular}{lc}
\toprule
(Branch 0, Branch 1) outcome & \# queries \\
\midrule
(good, good)                & 24 \\
(early\_stop, good)         & 14 \\
(good, early\_stop)         &  6 \\
(good, fail)                &  1 \\
(fail, good)                &  1 \\
(fail, fail)                & 23 \\
\bottomrule
\end{tabular}
\end{table}

Read together, the two distributions support the loop's design
choices. The $66.7\%$ acceptance rate is comparable to the strongest
reasoning-baseline Covered rate at $k\!=\!100$ on MathlibMPR,
indicating that the judge's binary feasibility decision is a
reasonable proxy for whether the underlying retrieval substrate has
actually surfaced the right premises. The fact that no accepted
query requires the maximum number of revisions also suggests the
revision cap is appropriately set: each additional revision past the
initial sketch yields fewer acceptances ($29 \to 11 \to 6$), so the
marginal yield of a fourth revision would be small. Finally, the
two branches rarely diverge in their final verdict --- $24 + 23 = 47$
queries see both branches finish in the same state, ignoring
\texttt{early\_stop} --- supporting the design choice of two
parallel branches as a low-cost variance-reduction mechanism rather
than a substantively different exploration strategy.

\subsection{Search task: per-slice breakdown}
\label{app:search-slices}

We report nDCG@$10$ and Recall@$10$ on three slicings of the fair
($810$-row) subset of \textbf{MathlibQR}: by query difficulty (Easy /
Hard), by ground-truth declaration kind (reporting the kinds for
which our encoder uses a kind-specific prompt template, alongside
the overall row), and by query style (six styles). Computing the
slices on the fair subset keeps the per-slice basis identical to
the main-report basis in Table~\ref{tab:search-main}. Throughout the
following tables we abbreviate column headers as
\textbf{LE} = LeanExplore, \textbf{LF} = LeanFinder,
\textbf{LSv2(r)} = LeanSearch~v2 retriever-only, and
\textbf{LSv2(rr)} = LeanSearch~v2 with reranking.

\begin{table}[ht]
\centering
\caption{\textbf{MathlibQR by difficulty.} Half the queries are Easy, half are Hard.}
\label{tab:search-by-difficulty}
\small
\setlength{\tabcolsep}{6pt}
\begin{tabular}{lcccccccc}
\toprule
& \multicolumn{4}{c}{nDCG@$10$} & \multicolumn{4}{c}{Recall@$10$} \\
\cmidrule(lr){2-5}\cmidrule(lr){6-9}
Difficulty & LE & LF & LSv2(r) & LSv2(rr) & LE & LF & LSv2(r) & LSv2(rr) \\
\midrule
Easy & 0.410 & 0.568 & 0.533 & \textbf{0.666} & 0.599 & 0.731 & 0.722 & \textbf{0.829} \\
Hard & 0.373 & 0.491 & 0.449 & \textbf{0.571} & 0.534 & 0.658 & 0.580 & \textbf{0.722} \\
\bottomrule
\end{tabular}
\end{table}

\begin{table}[ht]
\centering
\caption{\textbf{MathlibQR by ground-truth declaration kind on the fair subset.} We report the score on the kinds for which our encoder uses a kind-specific prompt template (theorem, def, instance) alongside the overall row, so that the per-kind cells are independent comparisons rather than an exhaustive partition. Best per row in \textbf{bold}.}
\label{tab:search-by-kind}
\small
\setlength{\tabcolsep}{5pt}
\begin{tabular}{lcccccccc}
\toprule
& \multicolumn{4}{c}{nDCG@$10$} & \multicolumn{4}{c}{Recall@$10$} \\
\cmidrule(lr){2-5}\cmidrule(lr){6-9}
Kind & LE & LF & LSv2(r) & LSv2(rr) & LE & LF & LSv2(r) & LSv2(rr) \\
\midrule
overall  & 0.393 & 0.533 & 0.494 & \textbf{0.623} & 0.569 & 0.698 & 0.657 & \textbf{0.780} \\
theorem  & 0.351 & 0.617 & 0.619 & \textbf{0.765} & 0.546 & 0.790 & 0.849 & \textbf{0.914} \\
def      & 0.441 & 0.637 & 0.657 & \textbf{0.764} & 0.571 & 0.790 & 0.829 & \textbf{0.943} \\
instance  & 0.390 & 0.661 & 0.839 & \textbf{0.846} & 0.619 & 0.866 & 0.907 & \textbf{0.938} \\
\bottomrule
\end{tabular}
\end{table}

\begin{table}[ht]
\centering
\caption{\textbf{MathlibQR by query style.} \textit{q1a} Lean-flavored phrasing; \textit{q1b} LaTeX statement; \textit{q1c} plain English statement; \textit{q2} short conceptual slogan; \textit{q3} informal nickname; \textit{q4} special-case instance.}
\label{tab:search-by-querytype}
\small
\setlength{\tabcolsep}{5pt}
\begin{tabular}{lcccccccc}
\toprule
& \multicolumn{4}{c}{nDCG@$10$} & \multicolumn{4}{c}{Recall@$10$} \\
\cmidrule(lr){2-5}\cmidrule(lr){6-9}
Query style & LE & LF & LSv2(r) & LSv2(rr) & LE & LF & LSv2(r) & LSv2(rr) \\
\midrule
q1a Lean        & 0.340 & 0.607 & 0.504 & \textbf{0.622} & 0.524 & \textbf{0.735} & 0.624 & 0.729 \\
q1b LaTeX      & 0.383 & 0.551 & 0.508 & \textbf{0.634} & 0.561 & 0.696 & 0.655 & \textbf{0.807} \\
q1c plain text  & 0.449 & 0.528 & 0.531 & \textbf{0.667} & 0.647 & 0.724 & 0.694 & \textbf{0.824} \\
q2 slogan       & 0.367 & 0.441 & 0.477 & \textbf{0.573} & 0.518 & 0.655 & 0.655 & \textbf{0.720} \\
q3 nickname     & 0.510 & 0.607 & 0.465 & \textbf{0.641} & 0.685 & 0.741 & 0.685 & \textbf{0.843} \\
q4 special case  & 0.079 & 0.218 & 0.317 & \textbf{0.500} & 0.217 & 0.348 & 0.522 & \textbf{0.783} \\
\bottomrule
\end{tabular}
\end{table}

LeanSearch~v2 with reranking leads on every reported slice $\times$ metric cell except a near-tie on q1a Lean queries, where LeanFinder edges out v2 by $0.006$ in Recall@$10$. The lead widens on the harder slices: on \emph{Hard} queries v2 with reranking is $+\,6.4$ Recall@$10$ ahead of the next-best system, and on \emph{q4 special case} --- the most adversarial slice for surface matching --- the lead is $+\,26.1$ Recall@$10$ over the next-best system in the table. The kind-sliced rows in Table~\ref{tab:search-by-kind} are reported as independent comparisons rather than a partition of the benchmark, since not every declaration kind admits a clean per-kind prompt template; an ablation that isolates the kind-aware instruction's contribution on the same three kinds is in Appendix~\ref{app:search-ablation}.

\subsection{Global premise retrieval: alternative success criteria}
\label{app:premise-ablation}

The headline numbers in Table~\ref{tab:premise-main} count a query as
solved when \emph{some} expert-annotated proof routing --- either the
original PR proof or one of the curator's equivalent alternative
routings --- has every premise group hit by the top-$k$. To verify that
the conclusions of \S\ref{sec:exp-premise} are not driven by the
allowance for alternative routings, we report a stricter Covered
variant that uses only the original PR routing, alongside the
main-text variant on the same set of $k$ values.
Table~\ref{tab:premise-solved-variants} reports both columns over
$k \in \{5, 10, 20, 30, 50\}$ for every system.

\begin{table}[ht]
\centering
\caption{\textbf{Covered variants on MathlibMPR ($n{=}69$, percentages).} \emph{Covered (main only)} requires the original PR routing's groups to all be hit; \emph{Covered (main $\vee$ alt-all)} additionally credits a query when an entire alternative routing's groups are hit, and is the variant reported in Table~\ref{tab:premise-main}. Best per column in \textbf{bold}.}
\label{tab:premise-solved-variants}
\small
\setlength{\tabcolsep}{4pt}
\begin{tabular}{lccccc@{\hspace{6pt}}ccccc}
\toprule
& \multicolumn{5}{c}{Covered (main only)} & \multicolumn{5}{c}{Covered (main $\vee$ alt-all)} \\
\cmidrule(lr){2-6}\cmidrule(lr){7-11}
System & 5 & 10 & 20 & 30 & 50 & 5 & 10 & 20 & 30 & 50 \\
\midrule
\multicolumn{11}{l}{\emph{Reasoning retrieval (informal $+$ formal query)}} \\
INF-X-Retriever                        & 14.5 & 17.4 & 18.8 & 21.7 & 21.7 & 15.9 & 18.8 & 21.7 & 24.6 & 24.6 \\
\quad $+$ query rewriter               & 10.1 & 10.1 & 14.5 & 15.9 & 23.2 & 10.1 & 10.1 & 17.4 & 17.4 & 24.6 \\
ReasonIR                               & 15.9 & 17.4 & 18.8 & 18.8 & 26.1 & 17.4 & 18.8 & 20.3 & 21.7 & 30.4 \\
DIVER (full pipeline)                  & 17.4 & 21.7 & 24.6 & 29.0 & 31.9 & 18.8 & 24.6 & 29.0 & 33.3 & 37.7 \\
\quad retriever only                   & 14.5 & 14.5 & 23.2 & 27.5 & 27.5 & 15.9 & 15.9 & 26.1 & 31.9 & 33.3 \\
\quad $+$ query expansion only         & 14.5 & 15.9 & 23.2 & 23.2 & 26.1 & 15.9 & 17.4 & 26.1 & 27.5 & 31.9 \\
\quad $+$ GroupRank rerank only        & 17.4 & 21.7 & 24.6 & 29.0 & 31.9 & 18.8 & 24.6 & 27.5 & 33.3 & 37.7 \\
\textbf{LeanSearch~v2 (reasoning)}     & \textbf{26.1} & \textbf{29.0} & \textbf{33.3} & \textbf{40.6} & \textbf{40.6} & \textbf{27.5} & \textbf{30.4} & \textbf{34.8} & \textbf{43.5} & \textbf{43.5} \\
\midrule
\multicolumn{11}{l}{\emph{Premise selection (proof-state query)}} \\
ReProver                               &  0.0 &  1.4 &  2.9 &  2.9 &  2.9 &  0.0 &  1.4 &  2.9 &  4.3 &  4.3 \\
LeanPremise                    &  1.4 &  1.4 &  4.3 &  4.3 &  5.8 &  1.4 &  1.4 &  4.3 &  4.3 &  5.8 \\
LeanStateSearch                        &  2.9 &  2.9 &  7.2 &  8.7 & 11.6 &  2.9 &  2.9 &  7.2 & 10.1 & 13.0 \\
\bottomrule
\end{tabular}
\end{table}

The relative ordering of systems is preserved between the two
variants: LeanSearch~v2 reasoning leads in every column, the DIVER
full pipeline remains the strongest non-v2 reasoning baseline, and
the gap between the strongest premise-selection system at any $k$
and v2 reasoning at $k\!=\!5$ is preserved under both criteria. The
absolute gap between Covered (main $\vee$ alt-all) and Covered
(main only) is on the order of $1$--$3$ percentage points for the
strong reasoning systems and is essentially zero for the
premise-selection systems, so the conclusions of \S\ref{sec:exp-premise}
do not depend on alternative-routing credit.

\subsection{MathlibQR across benchmark perspectives}
\label{app:mathlibqr-perspectives}

The main report uses the \textbf{fair} perspective on \textbf{MathlibQR}: every system can in principle return the ground-truth declaration because the query rows are restricted to the intersection of the three Mathlib snapshots (LeanFinder's, ours, and LeanExplore's). This is the cleanest comparison, but it discards $136$ rows whose ground-truth lives outside one or more competitor snapshots. To check that the conclusions of \S\ref{sec:exp-search} survive on looser comparators, we re-score the same retrievers under two additional perspectives. The \emph{at-least-one} perspective ($937$ rows) keeps every query whose ground-truth is in at least one of the three snapshots, and the \emph{full} perspective ($946$ rows) is the unfiltered benchmark, which includes nine rows whose ground-truth is in none of the three snapshots and so cannot be retrieved by any of the four systems.

\begin{table}[ht]
\centering
\caption{\textbf{LeanSearch~v2 vs.\ baselines across MathlibQR perspectives.} Each cell is nDCG@$k$ or Recall@$k$ on the listed perspective. Best per (perspective $\times$ column) cell in \textbf{bold}.}
\label{tab:perspectives}
\small
\setlength{\tabcolsep}{4pt}
\begin{tabular}{llcccccc}
\toprule
& & \multicolumn{3}{c}{nDCG@\,$k$} & \multicolumn{3}{c}{Recall@\,$k$} \\
\cmidrule(lr){3-5}\cmidrule(lr){6-8}
Perspective ($n$) & System & 1 & 5 & 10 & 10 & 50 & 100 \\
\midrule
\multirow{3}{*}{fair ($810$)}
  & LeanExplore                       & 0.246 & 0.358 & 0.393 & 0.569 & 0.743 & 0.789 \\
  & LeanFinder                        & 0.370 & 0.514 & 0.533 & 0.698 & 0.824 & 0.875 \\
  & \textbf{LeanSearch~v2 (rerank)}   & \textbf{0.470} & \textbf{0.601} & \textbf{0.623} & \textbf{0.780} & \textbf{0.847} & \textbf{0.858} \\
\midrule
\multirow{3}{*}{at-least-one ($937$)}
  & LeanExplore                       & 0.238 & 0.351 & 0.386 & 0.564 & 0.735 & 0.777 \\
  & LeanFinder                        & 0.320 & 0.445 & 0.461 & 0.603 & 0.712 & 0.757 \\
  & \textbf{LeanSearch~v2 (rerank)}   & \textbf{0.446} & \textbf{0.576} & \textbf{0.595} & \textbf{0.747} & \textbf{0.810} & \textbf{0.820} \\
\midrule
\multirow{3}{*}{full ($946$)}
  & LeanExplore                       & 0.236 & 0.347 & 0.383 & 0.558 & 0.728 & 0.770 \\
  & LeanFinder                        & 0.317 & 0.440 & 0.457 & 0.597 & 0.705 & 0.750 \\
  & \textbf{LeanSearch~v2 (rerank)}   & \textbf{0.442} & \textbf{0.570} & \textbf{0.590} & \textbf{0.740} & \textbf{0.802} & \textbf{0.812} \\
\bottomrule
\end{tabular}
\end{table}

LeanSearch~v2 with reranking ranks first on every (perspective $\times$ metric) cell. The lead over the second-best system is largest on the fair perspective ($+\,0.087$ nDCG@$5$ over LeanFinder, $+\,0.082$ Recall@$10$); on the looser perspectives the second-best slot shifts from LeanFinder to LeanExplore in some metrics, but our lead remains $+\,0.13$ nDCG@$5$ on at-least-one and $+\,0.13$ on full. The absolute scores fall $4$--$6$ points across columns when moving from fair to full, reflecting the extra $136$ rows that some systems cannot reach by construction; the relative ordering is unaffected.

\section{Ablation Study}

\subsection{Search-task ablations}
\label{app:search-ablation}

Standard mode is built up in two design steps: starting from a plain dense retriever, we first add a reranker, and then add a kind-aware instruction to both the corpus encoding and the reranker prompt. Both steps are intended to help, but the contributions are not symmetric and depend on the kind of declaration being retrieved. The ablation in Table~\ref{tab:search-progressive} isolates each step on the fair subset of \textbf{MathlibQR}, breaking the score down on the three declaration kinds for which our encoder uses a kind-specific prompt template (theorem, def, instance) and reporting Overall as the union.

\begin{table}[ht]
\centering
\caption{\textbf{Progressive ablation of standard mode on the fair subset of MathlibQR ($810$ rows).} \emph{no rerank}: Qwen3-Embedding-8B alone. \emph{rerank, no kind}: $+$ Qwen3-Reranker-8B with a kind-blind corpus encoding (every record encoded as \texttt{theorem}) and a kind-agnostic instruction. \emph{rerank, $+$kind} (production): kind-aware corpus encoding and instruction. Best per (slice $\times$ column) cell in \textbf{bold}.}
\label{tab:search-progressive}
\small
\setlength{\tabcolsep}{5pt}
\begin{tabular}{llccccc}
\toprule
Slice ($n$) & Pipeline & R@$1$ & R@$5$ & R@$10$ & nDCG@$5$ & nDCG@$10$ \\
\midrule
\multirow{3}{*}{Overall ($810$)}
  & no rerank                  & 0.340 & 0.586 & 0.657 & 0.472 & 0.494 \\
  & rerank, no kind            & 0.441 & 0.694 & 0.774 & 0.572 & 0.598 \\
  & \textbf{rerank, $+$kind}   & \textbf{0.470} & \textbf{0.714} & \textbf{0.780} & \textbf{0.601} & \textbf{0.623} \\
\midrule
\multirow{3}{*}{theorem ($291$)}
  & no rerank                  & 0.381 & 0.766 & 0.849 & 0.593 & 0.619 \\
  & rerank, no kind            & \textbf{0.612} & 0.856 & \textbf{0.914} & 0.746 & 0.765 \\
  & \textbf{rerank, $+$kind}   & \textbf{0.612} & \textbf{0.859} & \textbf{0.914} & \textbf{0.748} & \textbf{0.765} \\
\midrule
\multirow{3}{*}{def ($105$)}
  & no rerank                  & 0.495 & 0.752 & 0.829 & 0.633 & 0.657 \\
  & rerank, no kind            & 0.505 & \textbf{0.867} & 0.924 & 0.702 & 0.720 \\
  & \textbf{rerank, $+$kind}   & \textbf{0.581} & \textbf{0.867} & \textbf{0.943} & \textbf{0.738} & \textbf{0.764} \\
\midrule
\multirow{3}{*}{instance ($97$)}
  & no rerank                  & \textbf{0.773} & 0.887 & 0.907 & \textbf{0.833} & 0.839 \\
  & rerank, no kind            & 0.742 & 0.887 & 0.907 & 0.819 & 0.825 \\
  & \textbf{rerank, $+$kind}   & \textbf{0.773} & \textbf{0.897} & \textbf{0.938} & \textbf{0.833} & \textbf{0.846} \\
\bottomrule
\end{tabular}
\end{table}

The two design steps stack additively on every slice where headroom exists. On the Overall slice, nDCG@$5$ climbs from $0.472$ at the embedding-only baseline to $0.601$ at the production configuration, a total uplift of $+\,0.129$ points; the rerank step contributes about three quarters of that gain ($+\,0.101$) and the kind step the remaining quarter ($+\,0.029$). The breakdown shifts noticeably by kind. On theorem queries the rerank step alone delivers nearly the entire uplift, taking nDCG@$5$ from $0.593$ to $0.746$, while the kind step adds $\le\!0.002$ on top, indistinguishable from noise on the slice's $291$ queries; this is consistent with the encoder's default prompt template already being theorem-shaped, so explicitly tagging items as theorems contributes little. On def queries both steps contribute substantively (nDCG@$5$ $0.633 \to 0.702 \to 0.738$, with the rerank step supplying about two thirds and the kind step the remaining third), making this the clearest example of the two interventions stacking. On instance queries the embedding-only baseline is already at $0.833$ nDCG@$5$ with little headroom; the rerank-no-kind configuration dips by $0.014$ before the kind step pulls it back to parity at $0.833$, so on this slice the kind step's role is to prevent a small regression rather than to drive an improvement. Across all four slices, the production rerank-with-kind configuration is at least as good as either of the simpler variants on every metric, and never worse, so the two-step build-up is justified by the data on every reported cell.

\subsection{Simple reflection loop with a weaker prover backbone}
\label{app:prove-ablation}

Table~\ref{tab:prove-main} runs the simple reflection loop with Claude
Sonnet~4.5 as the prover. We ablate that choice by re-running the
same three retrieval modes --- no retrieval (L0), LeanSearch~v2
standard mode (L1), LeanSearch~v2 reasoning mode (the L2 \emph{Sketch}
variant) --- with Kimi~K2 Instruct in the prover role on FATE-H
($n\!=\!100$). All other roles in the loop (the query rewriter on
the standard-mode reflection path, the sketch generator / filter /
judge inside reasoning mode, and the REPL verifier) are kept
identical to the corresponding Sonnet rows of
Table~\ref{tab:prove-main}.

\begin{table}[ht]
\centering
\caption{\textbf{Prover-backbone ablation on FATE-H.} Each cell is the count of solved problems out of $100$ (percentages in parentheses). The Sonnet column reproduces the L0 / L1 / reasoning-mode rows of Table~\ref{tab:prove-main}; the Kimi column re-runs the same loops with Kimi K2 Instruct in the prover role only.}
\label{tab:prove-prover-ablation}
\small
\setlength{\tabcolsep}{10pt}
\begin{tabular}{lcc}
\toprule
Retrieval mode (prover input)            & Sonnet prover & Kimi prover \\
\midrule
No retrieval                             & 4 (4.0\%)     & 1 (1.0\%) \\
LeanSearch~v2 (standard mode)            & 14 (14.0\%)   & 8 (8.0\%) \\
\textbf{LeanSearch~v2 (reasoning mode)}  & \textbf{20 (20.0\%)} & \textbf{12 (12.0\%)} \\
\bottomrule
\end{tabular}
\end{table}

The relative ordering of the three retrieval modes is preserved when
the prover is downgraded from Sonnet to Kimi: reasoning mode ($12$)
$>$ standard mode ($8$) $>$ no retrieval ($1$), so the marginal
contribution of better retrieval information --- both at round one
(sketch and aggregated premises from reasoning mode) and at
reflection (LeanSearch hits via standard mode) --- carries through
to a weaker prover backbone. The absolute success rates drop by
roughly $40\%$ ($14 \to 8$, $20 \to 12$) when Sonnet is replaced by
Kimi, indicating that prover capability and retrieval capability
contribute roughly multiplicatively to end-to-end proof success
rather than acting as substitutes.

\subsection{Reflection rounds in reasoning mode}
\label{app:reflection-ablation}

The reasoning-mode loop in \S\ref{subsec:reasoning} caps revision at three rounds and uses the feasibility judge's accept/reject decision to decide when the sketch is ready for downstream consumption. To isolate the contribution of the judge-driven reflection step, we re-run the Prove task on FATE-H with reasoning mode configured to use zero revisions: the initial sketch is generated as usual, but the judge's verdict is ignored and the filtered premise set is passed straight to the prover regardless of whether the judge would have accepted it. Every other component of the loop (sketch generator, filter, prover, REPL verifier, hyperparameters) is held fixed at the production setting. Comparing the two runs measures the marginal contribution of the reflection step under a single-attempt budget.

\begin{table}[ht]
\centering
\caption{\textbf{Reflection-rounds ablation on FATE-H ($n{=}100$).} \emph{rfl $=0$}: the judge's verdict is ignored and the initial sketch's filtered premise set is passed straight to the prover. \emph{rfl $=3$}: production setting, the sketch reviser is invoked up to three times before the judge accepts or the loop falls back to the latest filtered output. Other reasoning-mode components are identical.}
\label{tab:reflection-ablation}
\small
\setlength{\tabcolsep}{10pt}
\begin{tabular}{lcc}
\toprule
Reasoning-mode reflection & \# solved & success rate \\
\midrule
rfl $=0$ (judge bypassed)              & 16 & 16.0\% \\
\textbf{rfl $=3$ (production)}         & \textbf{20} & \textbf{20.0\%} \\
\bottomrule
\end{tabular}
\end{table}

The full reflection loop adds $4$ solved problems on FATE-H over the no-reflection variant, a $+\,4$ percentage-point gain that confirms the qualitative design argument of \S\ref{subsec:reasoning}: the judge's structured rejection signal lets the sketch reviser repair sketches whose initial premise set the prover cannot complete. The gain is consistent with our reading of Appendix~\ref{app:reasoning-loop-stats}, where roughly one third of accepted queries needed at least one revision before the judge accepted, and is small enough on absolute terms that the ablation does not undermine the main-table claim --- reasoning mode wins on FATE-H even when reflection is disabled, retaining a $+\,2$-point lead over the strongest non-reasoning retriever (INF-X-Retriever at $16.0\%$ in Table~\ref{tab:prove-main}). Disabling reflection is therefore a meaningful but bounded loss; we keep it on by default because the additional cost is modest (Appendix~\ref{app:reasoning-loop-stats} reports a mean of one revision call per accepted query).

\section{Compute Resources and Licenses}

\subsection{Compute resources}
\label{app:compute}

GPU computation uses a small accelerator pool, and LLM-driven steps use the Claude Sonnet~4.5 and Kimi~K2 Instruct hosted APIs. For standard mode (\S\ref{sec:exp-search}), embedding the entire Mathlib corpus with Qwen3-Embedding-8B runs on a single accelerator and completes within approximately six hours, with reranking performed on demand for the per-query top-$50$; in downstream experiments standard mode is exposed as a service hosted on two accelerators, returning a ranked list in roughly $0.2$ seconds per query. For reasoning mode (\S\ref{sec:exp-premise}), running the full sketch--filter--judge--reviser loop on a single query costs approximately \$1.10\,USD at current API pricing. The remaining compute --- running the reasoning-retrieval baselines (INF-X, ReasonIR, DIVER and its sub-configurations), the premise-selection baselines on extracted proof states, and the per-slice and ablation runs reported in the appendix --- amounts to roughly three to four GPU-hours in total. 

\subsection{Licenses for existing assets}
\label{app:licenses}

\begin{itemize}
\item \textbf{Mathlib}~\citep{mathlib}: Apache~2.0 license. We use versions v4.28.0-rc1 and v4.29.1.
\item \textbf{Jixia}: Apache~2.0 license. \url{https://github.com/frenzymath/jixia}.
\item \textbf{Qwen3-32B}~\citep{qwen3}: Apache~2.0 license.
\item \textbf{Qwen3-Embedding-8B} and \textbf{Qwen3-Reranker-8B}~\citep{qwen3embedding}: Apache~2.0 license.
\item \textbf{Kimi K2 Instruct}~\citep{kimik2}: modified MIT license.
\item \textbf{Claude Sonnet 4.5}~\citep{sonnet45}: accessed via the Anthropic API under its published usage policy.
\item \textbf{Gemini 2.5 Pro}~\citep{gemini25}: accessed via the Google AI API under its published terms of service.
\item \textbf{FATE-H}~\citep{fate}: MIT License.
\end{itemize}

All baseline retrieval systems (LeanExplore, LeanFinder, ReProver, LeanPremise, LeanStateSearch, INF-X-Retriever, ReasonIR, DIVER) are cited in the main text and accessed via their published code or hosted APIs.

\end{document}